\documentclass[aps,prd,superscriptaddress,amsmath,groupedaddress,       
        reprint,showpacs,10pt]{revtex4-1}
\usepackage{graphicx}
\usepackage{revsymb}
\usepackage{dcolumn}
\newcommand{\msb}{\mathrm{\overline{MS}}}
\newcommand{\almsb}{\alpha_\msb}
\newcommand{\eq}[1]{Eq.~(\ref{#1})}
\newcommand{\order}{{\cal O}}

\newcommand{\psib}{\overline{\psi}}

\newcommand{\mcfit}{0.8905(56)}
\newcommand{\mcref}{0.9851(63)}
\newcommand{\mcmc}{1.2715(95)}
\newcommand{\alfit}{0.2128(25)}
\newcommand{\almz}{0.11822(74)}
\newcommand{\mcms}{11.652(65)}
\newcommand{\mslow}{93.6(8)}
\newcommand{\mshigh}{84.7(7)}
\newcommand{\mbms}{52.55(55)}
\newcommand{\mbmc}{4.528(54)}
\newcommand{\mbmb}{4.162(48)}
\newcommand{\knots}{\{2.9,  3.6,  4.6,  7.9\}}

\usepackage{times}

\begin{document}
	\title{High-precision quark masses and QCD coupling\\
	        from $n_f=4$ lattice QCD}
\author{Bipasha Chakraborty}
\affiliation{SUPA, School of Physics and Astronomy, University of Glasgow, Glasgow, G12 8QQ, UK}
\author{C.~T.~H.~Davies}
\affiliation{SUPA, School of Physics and Astronomy, University of Glasgow, Glasgow, G12 8QQ, UK}
\author{G.~C.~Donald}
\affiliation{Institut f\"{u}r Theoretische Physik, Universit\"{a}t Regensburg, 93040 Regensburg, Germany}
\author{R.~J.~Dowdall}
\affiliation{DAMTP, University of Cambridge, Wilberforce Road, Cambridge, CB3 0WA, UK}
\author{B.~Galloway}
\affiliation{SUPA, School of Physics and Astronomy, University of Glasgow, Glasgow, G12 8QQ, UK}
\author{P.~Knecht}
\affiliation{SUPA, School of Physics and Astronomy, University of Glasgow, Glasgow, G12 8QQ, UK}
\author{J.~Koponen}
\affiliation{SUPA, School of Physics and Astronomy, University of Glasgow, Glasgow, G12 8QQ, UK}
	\author{G.\ P.\ Lepage}
	\email{g.p.lepage@cornell.edu}
	\affiliation{Laboratory for Elementary-Particle Physics,
		Cornell University, Ithaca, NY 14853, USA}
\author{C. McNeile}
\affiliation{School of Computing and Mathematics 
and Centre for Mathematical Science,
Plymouth University, Plymouth PL4 8AA, United Kingdom}
	\collaboration{HPQCD Collaboration}
	\noaffiliation
	\date{4 December 2014}

\begin{abstract}
We present a new lattice QCD analysis of 
heavy-quark pseudoscalar-pseudoscalar correlators, using gluon 
configurations from the MILC collaboration 
that include vacuum polarization from $u$, $d$, $s$ and $c$~quarks
($n_f=4$). We extract new values for the QCD coupling and for 
the $c$ quark's $\msb$ mass: $\almsb(M_Z,n_f=5) = \almz$ and
$m_c(3\,\mathrm{GeV}, n_f=4) = \mcref$\,GeV. These agree well with our earlier
simulations using $n_f=3$ sea quarks, vindicating the perturbative treatment
of $c$ quarks in that analysis. 
We also
obtain a new nonperturbative result 
for the ratio of $c$~and $s$~quark
masses: $m_c/m_s=\mcms$. This ratio
implies $m_s(2\,\mathrm{GeV}, n_f=3)=\mslow$\,MeV when it is
combined with our new~$c$~mass. Combining
$m_c/m_s$ with our earlier $m_b/m_c$ gives 
$m_b/m_s=\mbms$, which is several standard deviations (but only 4\%)
away from the Georgi-Jarlskop prediction from certain GUTs. 
Finally we obtain an $n_f=4$ estimate for $m_b/m_c=\mbmc$ which 
agrees well with our earlier $n_f=3$ result. The new ratio 
implies~$m_b(m_b,n_f=5)=\mbmb$\,GeV.
\end{abstract}

\maketitle
\section{Introduction}
The precision of lattice QCD simulations has increased dramatically over 
the past decade, with many calculations now delivering 
results with 1--2\%~errors or less. Such precision requires increasingly 
accurate values for the fundamental QCD parameters: 
the quark masses and the QCD coupling. 
Accurate QCD parameters are important for non-QCD 
phenomenology as well. For example, theoretical uncertainties in several 
of the most important Higgs branching fractions are 
currently dominated by 
uncertainties in the heavy-quark masses (especially $m_b$ and $m_c$) and
the QCD coupling~\cite{*[{For a review see }][] Lepage:2014fla}.

In this paper we present new lattice results for~$m_c$,
$m_c/m_s$, $m_s$, $m_b/m_c$, $m_b$, and~$\alpha_s$. 
In a previous paper~\cite{McNeile:2010ji} we obtained 
0.6\%-accurate results for the masses and coupling by comparing continuum
perturbation theory with nonperturbative lattice-QCD 
evaluations of current-current correlators for heavy-quark currents. 
Current-current correlators are particularly well suited to a perturbative analysis
because non-perturbative effects are suppressed by four powers 
of $\Lambda_\mathrm{QCD}/2m_h$ where $m_h$ is the heavy-quark mass. 
Our earlier simulations treated $u$, $d$ and~$s$ sea quarks nonperturbatively ($n_f=3$), 
while assuming that contributions from $c$~and heavier quarks can be computed using 
perturbation theory. Here we test the assumption that heavy-quark contributions are 
perturbative by repeating our analysis with lattice simulations that treat the 
$c$~quark nonperturbatively ($n_f=4$ in the simulation).

In Section~2 we present our new $n_f=4$ lattice-QCD analysis of current-current
correlators, leading to new results for the heavy-quark masses and the QCD
coupling. We introduce an improved procedure that gives smaller errors
and simplifies the analysis.
We also demonstrate how our Monte Carlo data correctly reproduce the 
running of the $\msb$ masses and coupling.
In Section~3, we use the same simulations to calculate a new
nonperturbative result for the ratio of the $c$~to $s$~quark masses,~$m_c/m_s$. 
In Section~4, we use these simulations to calculate the mass ratio $m_h/m_c$ 
for heavy quarks with masses $m_h$ between $m_c$ and $m_b$. We express the
ratio as a function of the heavy quark's pseudoscalar mass~$m_{\eta_h}$. 
We extrapolate our result to $m_{\eta_h}=m_{\eta_b}$ 
to obtain a new nonperturbative 
estimate for~$m_b/m_c$.
In Section~5, we summarize our conclusions, derive new 
values for the $s$~and $b$~masses, and present our thoughts 
about further work in this area. 
We also include, in Appendix~\ref{app-msea},
a detailed discussion about how the coupling constant, quark masses, and 
the lattice spacing depend upon sea-quark masses in our approach. 
Our current analysis includes
 $u/d$~sea-quark masses down to physical values, 
so we are able to analyze this in far more detail than before.
Finally, Appendix~\ref{app-previous} briefly summarizes $n_f=4$ results obtained
using our previous methods~\cite{McNeile:2010ji}.

\section{Lattice Results}  

Our new analysis follows our earlier work~\cite{McNeile:2010ji}, but 
with a simpler and more accurate method for connecting current correlators
to $\msb$ masses. In particular, this method allows us to determine the
$\msb$ $c$~mass at multiple scales, from correlators with different heavy-quark
masses, providing a new test of our use of continuum perturbation theory. 
While the lattice spacings are not as small as before, 
our new analysis treats $c$-quarks in the quark sea 
nonperturbatively. We also use the substantially more accurate 
HISQ~discretization for the sea-quark action~\cite{Follana:2006rc}, in place of the 
ASQTAD~discretization in our earlier analysis, and a more accurate method
for setting the lattice spacing. The gluon action is also improved over our
earlier analysis, as it now includes~$\order(n_f \alpha_s a^2)$
corrections~\cite{Hart:2008sq}.
Our new results also have
more statistics, and include ensembles with $u/d$ masses very close
to the physical value. 

\subsection{Heavy-Quark Correlator Moments}
\label{sec:heavy-quark-correlators}
As before, we compute (temporal) moments
\begin{equation}
	G_n \equiv \sum_t (t/a)^n G(t)
\end{equation}
of correlators formed from the pseudoscalar density operator of a 
heavy quark, $j_5\equiv\overline{\psi}_h \gamma_5 \psi_h$:
\begin{equation}
	G(t) = a^6 \sum_{\mathbf{x}} (am_{0h})^2 \langle 0 | j_5(\mathbf{x},t)
	j_5(0,0) | 0 \rangle.
\end{equation}
Here $m_{0h}$ is the heavy quark's bare 
mass (from the lattice QCD lagrangian), $a$ is the lattice spacing, 
time~$t$ is Euclidean and periodic with period~$T$, and the sum
over spatial positions~$\mathbf{x}$ sets the total three-momentum to zero.
We again reduce finite-lattice spacing, tuning and perturbative errors by 
replacing the moments in our analysis with reduced moments:
\begin{equation}
	\tilde R_n \equiv \left\{
    \begin{aligned}
        & G_4/G^{(0)}_4 & & \text{for $n=4$,} \\
        &\frac{1}{m_{0c}}
            \left(G_n/G^{(0)}_n\right)^{1/(n-4)}
        & & \text{for $n\ge6$,}
    \end{aligned}\right.
	\label{rn-def}
\end{equation}
where $G^{(0)}_n$ is the moment in lowest-order weak-coupling perturbation 
theory using the lattice regulator, and
$m_{0c}$ is the bare mass of the $c$~quark. 

Low-$n$ moments are dominated by short-distance physics 
because the correlator is evaluated at 
zero total energy, which is well below the threshold for on-shell hadronic states: 
the threshold is at $E_\mathrm{threshold} = m_{\eta_h}$ where
\begin{equation}
	\mbox{2.9\,GeV} \le m_{\eta_h} < \mbox{6.6\,GeV}
\end{equation}
for our range of masses~$m_{0h}$. 
Furthermore, the moments are independent of the ultraviolet cutoff when $n\ge4$. 
Applying the Operator Product Expansion (OPE) to the product of 
currents in the correlator, we can therefore write our $n=4$ 
reduced moment in terms of
continuum quantities,
\begin{align}
	\label{rn-pert1}
	\tilde R_4 \to \,&r_4(\almsb,\mu) 
	\left\{ 1 +
	 \vphantom{
	 \sum_{q=u,d,s}\frac{\langle m_q \psib_q\psi_q\rangle}{(2m_h)^4}}
	\right. \nonumber \\
	& + 
	d_4^\mathrm{cond}(\almsb,\mu) 
	\frac{\langle\alpha_s G^2/\pi\rangle_\mathrm{eff}}{(2m_h)^4}	
	\nonumber\\
	& \left.
	+ \tilde{d}_4^\mathrm{cond}(\almsb,\mu) 
	\sum_{q=u,d,s}\frac{\langle m_q \psib_q\psi_q\rangle_\mathrm{eff}}{(2m_h)^4}
	+ \cdots
	\right\},
\end{align}
in the continuum limit ($a\to0$).
Here $\alpha_{\overline{\mathrm{MS}}}$ is the $\msb$ coupling at scale~$\mu$,
and $m_h$ is the $\msb$ $h$-quark mass. Heavy-quark condensates are 
absorbed into the gluon condensate~\cite{Shifman:1978bx}. 
We will retain terms only through 
the gluon condensate in what follows
since its contribution is already very small 
and contributions from other condensates will be much smaller.
We discuss the precise meaning 
of $\langle\alpha_s G^2/\pi\rangle_\mathrm{eff}$
below.
Reduced moments with $n\ge6$ can be written:
\begin{align}
	\label{rn-pert2}
    \tilde R_n \to \,&\frac{r_n(\almsb,\mu)}{m_c(\mu)}
	\left\{ 1   
	 \vphantom{\frac{\langle\alpha_s G^2/\pi\rangle}{(2m_h)^4}}
	\right. \nonumber \\
	& + \left.
	d_n^\mathrm{cond}(\almsb,\mu) 
	\frac{\langle\alpha_s G^2/\pi\rangle_\mathrm{eff}}{(2m_h)^4}
	+ \cdots
	\right\},
\end{align}
where $m_c(\mu)$ is the $\msb$~mass of the $c$~quark.
The continuum expressions for $\tilde R_n$ should agree with 
tuned lattice simulations
up to finite-lattice-spacing
errors of $\mathcal{O}((am_h)^2\alpha_s)$. The
perturbative expansions for the coefficient functions~$r_n$ are known 
through third order: see Table~\ref{tab:rn-pth}
and~\cite{Chetyrkin:2006xg,Boughezal:2006px,Maier:2008he,Kiyo:2009gb,Maier:2009fz}.
The expansions 
for~$d_n^\mathrm{cond}$ are known through first 
order~\cite{Broadhurst:1994qj}.

\begin{table}
\caption{Perturbation theory coefficients for $r_n$ with $n_f=4$ sea
quarks, where the heaviest sea quark has the same mass~$m_h$
 as the valence quark 
(that is, the quark used to make the currents in the current-current 
correlator).
Coefficients are defined by $r_n=1 + \sum_j r_{nj} \alpha_\msb^j(\mu)$
where $\mu = m_h(\mu)$. These coefficients are derived 
in~\cite{Chetyrkin:2006xg,Boughezal:2006px,Maier:2008he,Kiyo:2009gb,Maier:2009fz}.
}
\label{tab:rn-pth}
	\begin{ruledtabular}
		\begin{tabular}{rrrr}
		$n$ & $r_{n1}$ & $r_{n2}$ & $r_{n3}$ \\
		\hline
		 4 & $ 0.7427$ & $ 0.0088$ & $-0.0296$ \\ 
		 6 & $ 0.6160$ & $ 0.4976$ & $-0.0929$ \\ 
		 8 & $ 0.3164$ & $ 0.3485$ & $ 0.0233$ \\ 
		10 & $ 0.1861$ & $ 0.2681$ & $ 0.0817$  			
		\end{tabular}
	\end{ruledtabular}
\end{table}

Parameter~$\mu$ sets the scale for $m_c$ and for
$\almsb$ in~$r_n$. As in our previous paper, we take 
\begin{equation}
\mu = 3m_h(\mu) 
\end{equation} 
in order to 
improve the convergence of perturbation theory. In fact, however, 
our method is almost completely independent of the choice of~$\mu$, by design.
We can reexpress~$\mu$ in terms of the $\msb$ mass of the
$c$ quark,
\begin{equation}
\mu = 3 m_c(\mu) \frac{m_{0h}}{m_{0c}},
\end{equation}
since
ratios of quark masses are regulator independent: that is,  
\begin{equation}
\label{eq:mratios}
	\frac{m_{0h}}{m_{0c}} = \frac{m_h(\mu)}{m_c(\mu)}
\end{equation}
up to $a^2$ errors (for any~$\mu$).

Our reduced moments differ for~$n\ge 6$
from our earlier work: here we multiply by $1/m_{0c}$
in~\eq{rn-def} instead of $m_{\eta_h}/2m_{0h}$. The ratio
of $G$s in $\tilde R_{n\ge6}$ introduces a factor of $m_{0h}/m_h(\mu)$.
This becomes $1/m_c(\mu)$ when multiplied by $1/m_{0c}$ (by \eq{eq:mratios}). 
Consequently we can use moments calculated
with any heavy-quark mass~$m_{0h}$ to estimate
the $\msb$ $c$~mass (at $\mu=3m_h(\mu)$). Consistency 
among~$m_c$s coming from 
different $m_{0h}$~values is
an important test of the formalism. 

We could have used the bare mass of any quark, in place of
$m_{0c}$, in \eq{rn-def}. Then the $n\ge6$ moments would give
values for the $\msb$ mass of that quark. Alternatively we 
could leave the quark mass factor out, in which case these moments
give the factors~$Z_m(\mu)$ that convert any bare lattice quark mass
into the corresponding $\msb$ mass at scale~$\mu$. Heavy-quark 
current-current correlators, as used here,
provide an alternative to {RI-mom}~\cite{Martinelli:1994ty} 
and similar methods for determining both light and heavy quark masses.

The new definition for the reduced moments simplifies
our analysis since the variation of factor~$m_c(\mu)$ with $\mu$
is well known from perturbative QCD. The $m_{\eta_h}$ dependence of the
analogous factor 
($m_{\eta_h}/2m_h$) in the old analysis is unknown \emph{a priori}, and so
must be modeled in the fit.
We analyzed our data using the old definitions; the results, which 
agree with the results we find with the new methods, are
described briefly in Appendix~\ref{app-previous}.

\subsection{Lattice Simulations}
\label{sec:latsim}

\begin{table*}
\caption{Simulation parameters
for the gluon ensembles used in this paper~\cite{Bazavov:2010ru,Bazavov:2012xda},
with lattice spacings of approximately 0.15, 0.12, 0.09 and 0.06~fm,
and various combinations of sea-quark masses.
The parameters for each  simulation are: 
the inverse lattice spacing in units of $w_0=0.1715(9)$\,fm, 
the spatial~$L$ and temporal~$T$ lattice lengths, the number of 
gluon configurations~$N_\mathrm{cf}$ (each with multiple time
sources), the bare sea-quark masses 
in lattice units ($am_{0\ell}, am_{0s}, am_{0c}$), and 
the tuned bare $s$~and $c$~quark masses in~GeV.
The tuned $s$ and $c$~masses gives physical values for the $\eta_s$
and $\eta_c$ mesons, respectively. The $\ell$~mass is the 
average of the $u$~and $d$~masses, which are set equal in our 
simulations. $Z_m(\mu)$ is the ratio of the $\msb$ quark mass $m_q(\mu,n_f=4)$
to the corresponding bare (lattice) mass~$m_{0q}$ (see 
Section~\ref{sec:nf=4results}).
The last two entries for each ensemble indicate the
degree to which the sea-quark masses are detuned (see Appendix~\ref{app-msea}).
}
\label{tab:ensembles}
	\begin{ruledtabular}
	\begin{tabular}{clcccccccccrr}
	ensemble & $\quad w_0/a$ & $L/a$ & $T/a$ &  $N_\mathrm{cf}$ 
	&$a m_{0\ell}$ & $a m_{0s}$ & $a m_{0c}$ & 
	$m_{0s}^\mathrm{tuned}$ & $m_{0c}^\mathrm{tuned}$ & $Z_m(3\,\mathrm{GeV})$
	& $\delta m_{uds}^\mathrm{sea}/m_s$ & $\delta m_c^\mathrm{sea}/m_c$
	\\ \hline
 1&      1.1119(10)& 16& 48&      1020&   0.01300&   0.0650&   0.838&      0.0895(7)&       1.138(4)&       0.866(5)&      0.228(16)&      $-0$.058(8) \\
 2&       1.1272(7)& 24& 48&      1000&   0.00640&   0.0640&   0.828&      0.0890(7)&       1.130(4)&       0.872(6)&      0.046(14)&      $-0$.050(8) \\
 3&       1.1367(5)& 36& 48&      1000&   0.00235&   0.0647&   0.831&      0.0885(7)&       1.125(4)&       0.876(5)&     $-0$.048(13)&      $-0$.034(8) \\
\hline  4&      1.3826(11)& 24& 64&       300&   0.01020&   0.0509&   0.635&      0.0866(7)&       1.057(3)&       0.933(6)&      0.236(16)&      $-0$.044(8) \\
 5&       1.4029(9)& 32& 64&       300&   0.00507&   0.0507&   0.628&      0.0861(7)&       1.051(3)&       0.938(6)&      0.067(14)&      $-0$.035(8) \\
 6&       1.4149(6)& 48& 64&       200&   0.00184&   0.0507&   0.628&      0.0857(7)&       1.047(3)&       0.941(6)&     $-0$.040(13)&      $-0$.024(8) \\
\hline  7&      1.9330(20)& 48& 96&       300&   0.00363&   0.0363&   0.430&      0.0823(9)&       0.977(3)&       1.009(6)&      0.104(11)&      $-0$.021(8) \\
 8&       1.9518(7)& 64& 96&       304&   0.00120&   0.0363&   0.432&      0.0818(7)&       0.973(3)&       1.013(6)&     $-0$.011(13)&      $-0$.003(8) \\
\hline  9&      2.8960(60)& 48&144&       333&   0.00480&   0.0240&   0.286&      0.0778(7)&       0.912(3)&       1.080(7)&      0.365(19)&       0.045(9) \\
	\end{tabular}
	\end{ruledtabular}
\end{table*}

To extract the coupling constant and $c$~mass from simulations,
we use the simulations to compute nonperturbative values for the 
reduced moments $\tilde R_n$ with small $n\ge4$
and a range of heavy-quark masses~$m_{0h}$. We vary the lattice spacing, so
we can extrapolate to zero lattice spacing, and the sea-quark masses, so we
can tune the masses to their physical values. 

The gluon-field ensembles we use come from the MILC
collaboration and include $u$, $d$, $s$, and $c$ quarks in the quark 
sea~\cite{Bazavov:2010ru,Bazavov:2012xda}. 
The parameters that characterize these ensembles are given in 
Table~\ref{tab:ensembles}. The highly accurate HISQ 
discretization~\cite{Follana:2006rc} is 
used here for both the sea quarks 
and the heavy quarks in the currents used to create the 
correlators. This discretization was designed to minimize $(am_h)^2$ errors
for large $m_h$. Our previous work used HISQ quarks in the currents, but 
a less accurate discretization (ASQTAD) for the sea quarks.

We also quote tuned values 
for the bare $s$~and $c$~quark masses in Table~\ref{tab:ensembles}.
These are the quark masses that give the physical values for the 
$\eta_s$ and $\eta_c$ masses, as discussed 
in~Appendix~\ref{sec:tuning-masses}. This is the bare $c$~mass
we use in~\eq{rn-def} for~$\tilde R_n$.

\begin{table}
\caption{Simulations results for 
$\eta_h$ masses and reduced moments with various bare
heavy-quark masses~$am_{0h}$ and gluon ensembles (first column,
see Table~\ref{tab:ensembles}). 
Only data for $am_{0h}\le0.8$ are used in fits to the correlators.
}
\label{tab:metah-Rn}
	\begin{ruledtabular}
	\begin{tabular}{cclcccc}
& $am_{0h}$ & $\quad am_{\eta_h}$ & $\tilde R_4$ & $\tilde R_6$ & $\tilde R_8$ & $\tilde R_{10}$ 
\\ \hline
 1&   0.826&    2.22510(10)&      1.1627(1)&       0.937(3)&       0.885(3)&       0.856(3)\\
&   0.888&     2.33188(9)&      1.1477(1)&       0.937(3)&       0.893(3)&       0.867(3)\\
 2&   0.818&     2.21032(6)&      1.1643(0)&       0.943(3)&       0.890(3)&       0.860(3)\\
 3&   0.863&     2.28770(4)&      1.1528(0)&       0.947(3)&       0.900(3)&       0.872(3)\\
\hline  4&   0.645&    1.83976(11)&      1.1842(2)&       0.986(3)&       0.915(3)&       0.874(2)\\
&   0.663&    1.87456(12)&      1.1783(2)&       0.988(3)&       0.919(3)&       0.880(2)\\
 5&   0.627&     1.80318(8)&      1.1896(1)&       0.989(3)&       0.915(3)&       0.874(2)\\
&   0.650&     1.84797(8)&      1.1819(1)&       0.992(3)&       0.921(3)&       0.881(2)\\
&   0.800&     2.13055(7)&      1.1409(1)&       1.001(3)&       0.951(3)&       0.920(3)\\
 6&   0.637&     1.82225(5)&      1.1860(1)&       0.994(3)&       0.921(3)&       0.880(2)\\
\hline  7&   0.439&     1.34246(4)&      1.2134(1)&       1.013(3)&       0.921(3)&       0.877(2)\\
&   0.500&     1.47051(4)&      1.1886(1)&       1.029(3)&       0.946(3)&       0.903(3)\\
&   0.600&     1.67455(4)&      1.1565(1)&       1.048(3)&       0.978(3)&       0.939(3)\\
&   0.700&     1.87210(4)&      1.1315(0)&       1.059(3)&       1.002(3)&       0.968(3)\\
&   0.800&     2.06328(3)&      1.1118(0)&       1.064(3)&       1.019(3)&       0.991(3)\\
 8&   0.433&     1.32929(3)&      1.2160(1)&       1.015(3)&       0.922(3)&       0.877(2)\\
&   0.500&     1.47012(3)&      1.1885(0)&       1.033(3)&       0.950(3)&       0.906(2)\\
&   0.600&     1.67418(3)&      1.1564(0)&       1.052(3)&       0.982(3)&       0.943(3)\\
&   0.700&     1.87177(2)&      1.1315(0)&       1.063(3)&       1.006(3)&       0.972(3)\\
&   0.800&     2.06297(2)&      1.1117(0)&       1.068(3)&       1.023(3)&       0.995(3)\\
\hline  9&   0.269&     0.88525(5)&      1.2401(4)&       1.011(3)&       0.913(3)&       0.869(2)\\
&   0.274&     0.89669(5)&      1.2368(4)&       1.014(3)&       0.917(3)&       0.873(2)\\
&   0.400&     1.17560(5)&      1.1752(2)&       1.068(3)&       0.985(3)&       0.944(3)\\
&   0.500&     1.38750(4)&      1.1440(2)&       1.094(3)&       1.023(3)&       0.985(3)\\
&   0.600&     1.59311(4)&      1.1204(1)&       1.112(3)&       1.051(3)&       1.017(3)\\
&   0.700&     1.79313(4)&      1.1018(1)&       1.122(3)&       1.073(3)&       1.043(3)\\
&   0.800&     1.98751(3)&      1.0867(1)&       1.127(3)&       1.088(3)&       1.063(3)\\
&   0.900&     2.17582(3)&      1.0823(0)&       1.399(4)&       1.246(3)&       1.169(3)\\
&   1.000&     2.35773(3)&      1.0284(0)&       1.442(4)&       1.295(4)&       1.215(3)\\
	\end{tabular}
	\end{ruledtabular}
\end{table}

In Table~\ref{tab:metah-Rn} we list our simulation results 
for the $\eta_h$ mass and the reduced moments for 
various bare quark masses $am_{0h}$ on various ensembles. 
Results from different values of $am_{0h}$ on the same ensemble
are correlated; we include these correlations in our analysis.
The $am_{\eta_h}$ values are computed from 
Bayesian fits of multi-state function
\begin{equation}
	\sum_{j=1}^{10} b_j \left(e^{-m_j t} + e^{-m_j(T-t)}\right)
\end{equation}
to the correlators~$G(t)$ for $t\ge8$, where $T$ is the temporal length of 
the lattice~\cite{Lepage:2001ym}. 
The fitting errors are small for $am_{\eta_h}$ and have minimal
impact on our final results.

The fractional errors in the $\tilde R_n$ for $n\ge6$ are 20--40~times
larger than those for~$\tilde R_4$. This is because of the factor
of $1/m_{0c}^\mathrm{tuned}$ used in~\eq{rn-def}
to define these moments. As mentioned above, 
we could have used bare masses for other quarks
in this definition, to obtain values for their $\msb$~masses.
Heavy-quark masses like $m_{0c}$, however, 
can usually be tuned more accurately than light-quark masses, as
discussed in Appendix~\ref{app-msea}. Masses for other quarks
can be obtained from the $c$~mass and nonperturbatively determined
quark mass ratios, as we show for the $s$~and $b$~masses in the next 
two sections.

As in our previous paper, we limit the maximum size of~$am_h$
in our analysis: we require~$am_h\le0.8$. 
This keeps $a^2$~errors smaller than~10\%.

We determine the lattice spacing by measuring the 
Wilson flow parameter $w_0/a$ on the lattice 
(Table~\ref{tab:ensembles})~\cite{Borsanyi:2012zs}. 
From previous simulations~\cite{Dowdall:2013rya}, we know that
\begin{equation}
	w_0 = 0.1715(9)\,\mathrm{fm},
\end{equation}
which we combine with our measured values of $w_0/a$ to obtain the lattice
spacing for each ensemble (Appendix~\ref{app-msea}). 
This approach is far more accurate than
that used in our earlier paper, which relied upon the $r_1$~parameter from
the static-quark potential.

\subsection{Fitting Lattice Data}
Our goal is to find values for $\almsb(\mu)$ and $m_c(\mu)$ 
that make the theoretical results (from perturbation theory) for the reduced 
moments $\tilde R_n$ (Eqs.~(\ref{rn-pert1}--\ref{rn-pert2})) agree with the 
nonperturbative results 
from our simulations. We do this by simultaneously fitting results from all
of our lattice spacings and quark masses for moments with $4\le n\le10$.
To get good fits, we must correct the continuum formulas in 
Eqs.~(\ref{rn-pert1}--\ref{rn-pert2}) for several systematic errors
in the simulation. We fit the lattice data using 
the following corrected form:
\begin{align}
	\label{eq:Rn-begin}
	\tilde R_n &= 
	\left.
	\begin{cases}
		1 & \mbox{for $n=4$} \\
		1/\xi_m m_c(\xi_\alpha \mu)
		& \mbox{for $n\ge6$}
	\end{cases}
	\right\}	
	\\
	& \times r_n(\almsb(\xi_\alpha\mu),\mu) 	
	\label{eq:Rn-2}
	\\
	& \times\left( 1 +
	d_n^\mathrm{cond} \frac{\langle\alpha_s G^2/\pi\rangle}{(2m_h)^4}
	\right) 
	\label{eq:Rn-3}
	\\
	& \times
		\label{eq:Rn-4}
	\left(1 + d_n^{h,c}\frac{m_{0h}^2-m_{0c}^2}{m_{0h}^2}\right) \\
	& + 
	\left(\frac{am_{\eta_h}}{2.26}\right)^2
	\sum_{i=0}^{N}
	c_{i}(m_{\eta_h},n) \left(\frac{am_{\eta_h}}{2.26}\right)^{2i}.
	\label{eq:Rn-end}
\end{align}

We use a Bayesian fit with priors for every fit parameter~\cite{Lepage:2001ym}.
The priors are \emph{a priori} estimates for the parameters based upon
theoretical expectations and previous experience, especially from our 
earlier, very similar $n_f=3$ analysis. In each case we test our 
choice of prior width against the Empirical Bayes criterion~\cite{Lepage:2001ym},
which in effect uses fluctuations in the data to suggest natural widths 
for priors. None of our priors is narrower than this optimal width,
and most are wider, which leads to more conservative
errors.

We now explain each part of the lattice formula in turn.

\subsubsection{Detuned Sea-quark Masses}
The terms $\almsb(\xi_\alpha\mu)$ and $\xi_m m_h(\xi_\alpha \mu)$  
in $\tilde R_n$ are the 
$\msb$  coupling and heavy-quark mass for detuned sea-quark masses;
see Eqs.~(\ref{eq:alpha-tilde}) and~(\ref{eq:m-tilde}) in 
Appendix~\ref{app-msea}.  Scale $\mu$ is chosen so that
\begin{equation}
	\mu = 3\, \xi_m m_c(\xi_\alpha\mu) 
	\frac{m_{0h}}{m_{0c}} = 3\,m_h(\mu,\delta m^\mathrm{sea}).
		\label{eq:mu-tilde}
\end{equation}

Scale factors $\xi_\alpha$ and $\xi_m$ are defined in Appendix~\ref{app-msea},
which discusses how $\msb$ couplings and masses are affected by sea-quark masses.
The coefficients $g_\alpha$, $g_m$\,\ldots in $\xi_\alpha$ and $\xi_m$ are 
treated as fit parameters, with priors taken from the output of the 
fits described in the appendix.

The light sea-quark masses enter linearly in $\xi_\alpha$ and $\xi_m$,
because of (nonperturbative) chiral symmetry breaking. 
Quark mass dependence also enters through the perturbation theory for 
the moments ($r_n$), but is quadratic in the mass 
and therefore negligible for light quarks.

\subsubsection{$\mu$ Dependence}
The scale $\mu$ enters Eqs.~(\ref{eq:Rn-begin})--(\ref{eq:Rn-end})
through the coupling constant $\almsb(\xi_\alpha\mu)$ and the
$c$~mass~$m_c(\xi_\alpha\mu)$. We parameterize the coupling and mass in the fit
by specifying their values at $\mu=5$\,GeV with fit parameters~$\alpha_0$
and $m_0$,
\begin{align}
	\label{eq:alpha0}
	\almsb(5\,\mathrm{GeV}, n_f=4) = \alpha_0 \nonumber \\
	m_c(5\,\mathrm{GeV}, n_f=4) = m_0,
\end{align}
whose priors are
\begin{align}
	\alpha_0 = 0.21 \pm 0.02, \quad
	m_0 = 0.90 \pm 0.10.
\end{align}
Our previous analysis gave $0.2134(24)$ and $0.8911(56)$ for these
parameters, so the priors are broad.
The coupling and mass for other values of~$\mu$ are obtained
by integrating (numerically) their evolution equations from 
perturbative QCD, starting from the values at~$\mu=5$\,GeV:
\begin{align}
	\mu^2 \frac{d\almsb(\mu)}{d\mu^2} 
	= &-\beta_0 \almsb^2(\mu)  - \beta_1 \almsb^3
	- \beta_2 \almsb^4 \nonumber \\
	&- \beta_3 \almsb^5 - \beta_4\almsb^6,
	\label{eq:almsb-evolution}
	\\
	\frac{d\log m_h(\mu)}{d\log \mu^2} =
	&- \gamma_0 \almsb(\mu) - \gamma_1 \almsb^2 - \gamma_2 \almsb^3 \nonumber\\
	&- \gamma_3 \almsb^4 - \gamma_4 \almsb^5.
	\label{eq:mh-evolution}	
\end{align}
The first four coefficients on the right-hand-sides of these equations are
known from perturbation 
theory~\cite{vanRitbergen:1997va,Czakon:2004bu,Vermaseren:1997fq,Chetyrkin:1997dh}. 
In each case, we treat the fifth coefficient as 
a fit parameter whose prior's width equals the root-mean-square average of 
the first four parameters:
\begin{align}
	\beta_4 = 0 \pm \sigma_\beta, \quad 
	\gamma_4 = 0 \pm \sigma_\gamma.
\end{align}
Neither $\beta_4$ nor $\gamma_4$ has signficant impact on our final results.

\subsubsection{Truncated Perturbation Theory}
\label{sec:truncated-pth}
The Wilson coefficient function~$r_n$ (\eq{eq:Rn-2}) has a 
perturbative expansion of the 
form
\begin{equation}
\label{eq:rn}
		r_n(\almsb,\mu) 
		\equiv 1 + \sum_{j=1}^{N_\mathrm{pth}} 
	r_{nj}(\mu) \almsb^j.
\end{equation}
The perturbative coefficients~$r_{nj}$ are known through third order,
and are given for $\mu=m_h(\mu)$ in Table~\ref{tab:rn-pth}.

The lack of perturbative coefficients beyond third order is our largest 
single source of systematic error. Our data are 
sufficiently precise that higher-order terms are relevant. Furthermore
the relative importance of the higher-order terms varies with quark mass, 
as~$\almsb$ varies with~$\mu=3m_h$. Therefore we 
include the higher-order terms in our analysis with coefficients that we fit
to account for variations with quark mass.
As in our earlier analysis, we note that
the known perturbative coefficients are small  
and relatively uncorrelated from moment to moment
and order to order
for $\mu=m_h$, leading us to adopt fit priors
\begin{equation}
\label{eq:rnprior}
	r_{nj}(\mu=m_h) = 0 \pm 1
\end{equation}
for the $n>3$ coefficients at $\mu=m_h$. We double the width of 
these priors relative to our previous analysis because the fit
suggested that some higher-order coefficients are larger here
(especially for $n=4$).

We set $N_\mathrm{pth}=15$ terms in the expansion, 
although our results are essentially unchanged once 8~or more terms are
included (or~5 with $\mu=m_h$). 
As before we use renormalization group equations to express
the coefficients $r_{nj}(\mu=3m_h)$ in terms of the 
coefficients~$r_{nj}(\mu=m_h)$ from Table~\ref{tab:rn-pth} and~\eq{eq:rnprior}. 
This procedure generates (correlated) priors for the
unknown coefficients at $\mu=3m_h$ that account for renormalization-group
logarithms. The procedure makes our results largely independent of~$\mu$:
our results change by less than a third of a standard deviation as 
$\mu$ is varied over the interval~$2m_h\le\mu\le 10 m_h$.

\subsubsection{Nonperturbative Effects; Finite-Volume Corrections}
We use the Operator Product Expansion (OPE) in 
Eqs.~(\ref{rn-pert1}--\ref{rn-pert2}) to separate short-distance 
from long-distance 
physics. In principle, the perturbative coefficients 
in $r_n(\almsb,\mu)$ above should
have subtractions coming from the higher-order terms in the OPE 
expansion:
\begin{equation}
\label{eq:rn-subtraction}
	r_n \to r_n \left(1 - d_n^{\mathrm{cond}} 
	\frac{\langle \alpha_s G^2/\pi\rangle^{(\lambda)}_\mathrm{pth}}{(2m_h)^4}
	- \cdots
	\right)
\end{equation}
where $\lambda$ is a fixed cutoff 
scale in the perturbative regime, say $\lambda=1$\,GeV, and 
$\langle \alpha_s G^2/\pi\rangle^{(\lambda)}_\mathrm{pth}$ 
and $d_n^\mathrm{cond}$ are
computed in perturbation theory to the same order as $r_n$. These 
subtractions come from perturbative matching, 
and remove contributions to $r_n$ 
due to 
low-momentum gluons ($q\!\le\!\lambda$), thereby also removing infrared 
renormalons order-by-order in perturbation theory.
The size of the subtraction depends
upon the detailed definition of $\alpha_s (G^{(\lambda)})^2$. This 
procedure is completely unambiguous given a specific definition
for this operator, but we have not
included the subtraction in $r_n$ since it is negligible for any reasonable 
definition at our low orders 
of perturbation theory. For example, 
a simple momentum-space cutoff, that keeps $q^2<\lambda^2$,
gives~\cite{Novikov:1984rf}
\begin{equation}
\label{eq:pert-G2}
	\langle \alpha_s G^2 \rangle^{(\lambda)}_\mathrm{pth} =
	\frac{3\alpha_s}{2\pi^3}\,\lambda^4,
\end{equation}
which ranges from 0.001 to 0.019\,$\mathrm{GeV}^4$ 
for $\lambda$s between 500\,Mev and 1\,GeV. This would change $r_n$
by no more than 0.1--0.4\% at $m_h=m_c$ and much less at 
our higher~$m_h$s.

Not surprisingly, perturbative estimates of the condensate value (\eq{eq:pert-G2})
are similar in size to nonperturbative estimates
of the condensate value. So it is simpler for us to combine the 
subtraction in \eq{eq:rn-subtraction} 
with the condensate itself to form an effective condensate  
value~\cite{Shifman:1998rb}:
\begin{equation}
	\langle \alpha_s G^2 \rangle_\mathrm{eff}
	\equiv \langle \alpha_s G^2 \rangle^{(\lambda)} 
	- \langle \alpha_s G^2 \rangle^{(\lambda)}_\mathrm{pth}
\end{equation}
In our fits we take $\langle \alpha_s G^2 \rangle_\mathrm{eff}$ as a fit
parameter with prior
\begin{equation}
	\label{eq:cond}
	\langle \alpha_s G^2 \rangle_\mathrm{eff} = 0.0 \pm 0.012,
\end{equation}
and we approximate $m_h\approx m_{\eta_h}/2.26$ in the condensate
correction (because $m_b(m_b)\approx m_{\eta_b}/2.26$). 
Our results are completely unchanged if the width
of this prior is ten times larger. In either case we obtain a value for
the effective condensate of order~$0.002$ with errors of a similar size. 
This is completely consistent 
with expectations, and it reduces condensate contributions to the 
moments to 0.01--0.05\% at~$m_h=m_c$, and much less at 
higher~$m_h$\,---\,negligible at our level of precision.

This procedure is sensible at our level of precision. As precision
increases, however, there is a point where it becomes important to remove 
renormalon corrections from the coefficients in $r_n$. Otherwise $j!$~factors
in $j^\mathrm{th}$~order, coming from infrared renormalons,
cause perturbation theory to diverge. A simple 
analysis~\cite{*[{See, for example, }][] Shifman:2013uka}
indicates that 
perturbation theory starts to diverge at order $j\sim 2/(\beta_0\almsb)$, 
which is around $8^\mathrm{th}$~order for our analysis. Consequently
we expect the impact of infrared
renormalons to be negligible at $3^\mathrm{rd}$~order.

Perturbation theory is not the whole story even if infrared renormalons
are removed. 
The OPE separates short-distances from long-distances, but the 
short-distance coefficients $r_n$, $d_n^\mathrm{cond}$\,\ldots\,have
nonperturbative contributions, for example, from small 
instantons~\cite{Novikov:1984rf}. 
It is also possible that the OPE is an
asymptotic expansion and does not
converge ultimately, although recent results suggest it 
might converge~\cite{Hollands:2011gf,Pappadopulo:2012jk}. 
Whatever the case, such effects are expected to appear at even
higher orders
than infrared renormalons, and so are completely negligible at our
level of precision. 

Condensates, renormalons, small instantons, 
\emph{etc.}\ afflict all perturbative analyses 
at some level of precision. Our analysis is particularly insensitive to 
such effects because the leading nonperturbative contributions are suppressed
by four powers of $\Lambda_\mathrm{QCD}/(2m_h)$.

Note finally that the coefficient functions, being short-distance, are
insensitive to errors caused by the finite volume of the lattice. 
While the finite
volume \emph{can} 
affect the value of $\langle \alpha_s G^2 \rangle_\mathrm{eff}$,
the impact on our results is negligible 
since the condensate itself is negligible. We verified this 
by recalculating the reduced moments for emsemble~5 in Table~\ref{tab:ensembles}
with spatial lattice sizes of $L/a=24$ and~40 (ensemble~5 uses~32). 
The moments for different volumes agree to within statistical
errors of order~0.01\%. The same is true for the measured values
of $m_{\eta_c}$ from these ensembles; finite volume effects 
will be smaller still for~$m_{\eta_h}$.

\subsubsection{$m_{0h}-m_{0c}$ Correction}
Our results are also affected by the difference between the $c$~mass 
$m_{0c}$ used in the sea, and the mass of the heavy quark $m_{0h}$ used 
to make the currents in the current-current correlator. The perturbative
calculations we use
assume $m_{0c}=m_{0h}$, but we want to study a range of $m_{0h}$ values with 
fixed $m_{0c}$. The correction enters in~$\order(\alpha_s^2)$,
is quadratic in the mass difference 
for small differences, and goes to a (small) constant as $m_{0h}\to\infty$. 
Therefore we correct for it using (\eq{eq:Rn-4})
\begin{equation}
 	\tilde R_n \to \tilde R_n \left(1 
 	+ d_n^{h,c} \frac{m_{0h}^2-m_{0c}^2}{m_{0h}^2}
 	\right)
 \end{equation} 
where $h_n$ is a fit parameter with a prior of $0\pm0.03$. The width~0.03 
is ten times larger than the correct value (from perturbation theory) 
in the $m_{0h}\to\infty$ limit. It is twice as wide as the width indicated
by the Empirical Bayes criterion~\cite{Lepage:2001ym}. We also tried fits
where $d_n^{h,c}$ was replaced by a spline function of $m_{\eta_h}$. These give
similar results but with larger errors (especially for~$\almsb$).

\subsubsection{Finite Lattice Spacing Errors}
The final modification in our formula for $\tilde R_n$ 
corrects for errors caused by the finite lattice spacings used in the 
simulations. We write 
\begin{equation}
	\tilde R_n \to \tilde R_n + \delta \tilde R_n
\end{equation}
where
\begin{equation}
	\delta \tilde R_n \equiv \left(\frac{am_{\eta_h}}{2.26}\right)^2
	\sum_{i=0}^{N}
	c_{i}^{(n)}(m_{\eta_h}) \left(\frac{am_{\eta_h}}{2.26}\right)^{2i}
\end{equation}
and again $m_{\eta_h}/2.26$ is a proxy for the quark mass.
We parameterize the $m_{\eta_h}$~dependence 
of the~$c_i^{(n)}(m_{\eta_h})$ using cubic splines with knots,
at
\begin{equation}
\label{eq:knots}
m_\mathrm{knots} \equiv \mbox{\knots\,GeV},
\end{equation}
that come from the analysis in Section~\ref{sec:mh/mc}.
 We set
\begin{equation}
	c_i^{(n)}(m) = c_{0i}^{(n)} + \delta c_i^{(n)}(m)
\end{equation}
with the following fit parameters and priors:
\begin{align}
		c_{0i}^{(n)} &= 0 \pm 1/n \nonumber \\
	\delta c_i^{(n)}(m) &= 0 \pm 0.10/n & & m\in m_\mathrm{knots}
	\nonumber \\
	\delta c_i^{(n)\prime}(m) &= 0 \pm 0.10/n & &m=2.9\,\mathrm{GeV}.
\end{align}
These priors are again conservative since the Empirical Bayes 
criterion~\cite{Lepage:2001ym} suggests priors that are half as wide.
We take $N=20$ but our results are insensitive to any~$N\ge10$.

\subsection{$n_f=4$ Lattice Results}
\label{sec:nf=4results}
We fit all of the reduced moments from our simulation data\,---\,with 
lattice spacings from 0.12\,fm to 0.06\,fm, and $n=4$, 6, 8 
and~10 in Table~\ref{tab:metah-Rn}\,---\,simultaneously to 
formula~(\ref{eq:Rn-begin}--\ref{eq:Rn-end}) by 
adjusting fit parameters described in the previous sections. 
The fit is excellent with a $\chi^2$~per degree of freedom of~0.51 for 
92~pieces of data ($p$-value is 1.0).

The fit has two key physics outputs. One is a new result for the 
running coupling constant:
\begin{equation}
	\almsb(5\,\mathrm{GeV}, n_f=4) = \alfit.
\end{equation}
To compare with our old determination and other determinations, we
use perturbation theory to add $b$~quarks to the sea~\cite{Chetyrkin:1997un}, 
with $m_b(m_b)=4.164(23)$\,GeV~\cite{McNeile:2010ji},
and evolve to the $Z$~mass (91.19\,GeV) to get
\begin{equation}
		\almsb(M_Z, n_f=5) = \almz.
\end{equation}
This agrees well with~$0.1183(7)$ from our 
$n_f=3$ analysis~\cite{McNeile:2010ji}.
It also agrees well with the current world average~0.1185(6) from the 
Particle Data 
Group~\cite{*[][{ and 2013 partial update for the 2014 edition.}] Agashe:2014kda}.

The second important physics output is the $c$~quark's mass, whose
value at $\mu=5$\,GeV is a fit parameter:
\begin{equation}
	m_c(\mu,n_f=4) = 
	\begin{cases}
		\mcfit\,\mathrm{GeV} & \mu=5\,\mathrm{GeV} \\
		\mcref\,\mathrm{GeV} & \mu=3\,\mathrm{GeV} \\
		\mcmc\,\mathrm{GeV} & \mu=m_c(\mu),
	\end{cases}
\end{equation}
where we have used \eq{eq:mh-evolution} to evolve our result to other 
scales for comparison with other determinations. These again agree 
well with our previous~$n_f=3$ analysis~\cite{McNeile:2010ji}, 
which gave~0.986(6)\,GeV for the mass at~3\,GeV. The errors 
for~$m_c(3\,\mathrm{GeV})$ and~$\almsb(M_Z)$ are correlated, 
with correlation coefficient~0.19.

We use our result from $m_c$ to calculate the mass renormalization factors
\begin{equation}
	Z_m(\mu) \equiv \frac{m_c(\mu)}{m_{0c}}
\end{equation}
that relate $\msb$ masses to bare lattice masses for each configuration. 
These factors can be used to convert the bare mass for any quark to its
$\msb$ equivalent. We tabulate these results, with $\mu=3$\,GeV, 
for our configurations in
Table~\ref{tab:ensembles}. 
These $Z_m$~values are much more accurate than can be obtained 
from order~$\alpha_s$ lattice QCD perturbation theory~\cite{McNeile:2012xh},
but they agree qualitatively and suggest that higher-order corrections 
from lattice perturbation theory are small. 

Our results confirm that a 
perturbative treatment of $c$~quarks in the sea,
as in our previous paper, is correct, at least to our current 
level of precision. 

Our result at $\mu=m_c$ has a larger error because $\almsb$ in the
mass evolution equation (\eq{eq:mh-evolution}) becomes fairly
large at that scale ($\almsb\approx\!0.4$) and quite sensitive to 
uncertainties in its value. We use the coupling from our fit for this
evolution.
Were we instead to use the Particle Data Group's (more accurate)~$\almsb$, 
our value for $m_c(m_c)$ would be
\begin{equation}
	m_c(m_c, n_f=4) = 1.2733(76)\,\mathrm{GeV}.
\end{equation}
In any case, it is probably better to 
avoid such low scales, if possible.

\begin{figure}
	\includegraphics[scale=0.9]{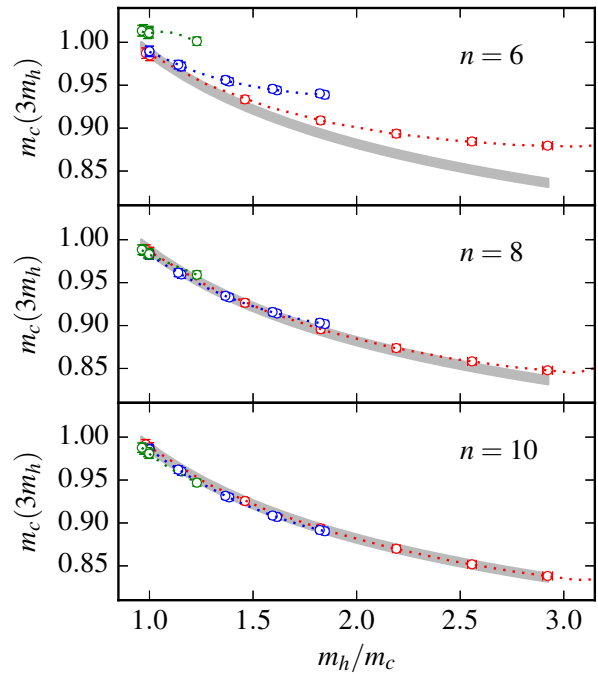}
	\caption{The $c$~quark mass $m_c(\mu=3m_h)$ as determined from moments
	with heavy-quark masses ranging from $m_c$ to~$2.9\,m_c$. The data points
	show results obtained by substituting nonperturbative simulation values 
	for $\tilde R_n$ into \eq{eq:nonpert-mc}, after correcting
	for mistunings of the sea-quark masses 
	(using the fit). Errors are about the size of the plot symbols, or smaller.
	Results are shown for 
	three lattices spacings: 0.12\,fm (green points, through
	$m_h/m_c=1.2$), 0.09\,fm (blue points, through $m_h/m_c=1.8$),
	and 0.06\,fm (red points, through $m_h/m_c=2.9$). The dotted
	lines show our fits to these data points. The gray band shows the
	values expected from our best-value $m_c(5\,\mathrm{GeV})=\mcfit$\,GeV evolved
	perturbatively to the other scales.
	}
	\label{fig:mc-mh}
\end{figure}

Note that our $c$~mass comes from moments whose heavy-quark mass varies
from $m_h=m_c$ to $m_h=3m_c$. Each (nonperturbative)~$\tilde R_n$ 
with $n\ge6$, for each heavy-quark mass~$m_h$, 
gives an independent estimate of the $c$~mass:
\begin{equation}
\label{eq:nonpert-mc}
	m_c(3m_h) = 
	\frac{r_n(\almsb(3m_h),\mu=3m_h)}{\tilde R_n}.
\end{equation}
The extent to which these estimates agree with each other is shown in 
Figure~\ref{fig:mc-mh}, where the nonperturbative results (data points) 
are compared with our best-fit result for $m_c(5\,\mathrm{GeV})$ evolved 
perturbatively to other scales using~\eq{eq:mh-evolution} (gray band).
As expected, 
finite~$a^2$ errors are larger for smaller values of~$n$
and larger values of~$m_h$~\cite{Allison:2008xk,McNeile:2010ji}. 
Taking account of these errors,
agreement between different determinations of the mass is excellent.

\begin{table}
	\caption{Error 
	budget~\cite{*[{The precise definition of our error budgets is described
	in Appendix~A of }][] Bouchard:2014ypa} 
	for the $c$~mass, QCD coupling,
	and the ratios of quark masses $m_c/m_s$ 
	and $m_b/m_c$ from the $n_f=4$ simulations
	described in this paper. Each uncertainty is given as a percentage
	of the final value. The different uncertainties are added in 
	quadrature to give the total uncertainty. Only sources of 
	uncertainty larger than~0.05\% have been listed.
	}
	\label{tab:nf4errors}
	\begin{ruledtabular}
	\begin{tabular}{rcccc}
		&$m_c(3)$ & $\almsb(M_Z)$ & $m_c/m_s$ & $m_b/m_c$
		\\ \hline
		Perturbation theory 							& 0.3 & 0.5 & 0.0 & 0.0 \\
		Statistical errors 								& 0.2 & 0.2 & 0.3 & 0.3 \\
		$a^2\to 0$               						& 0.3 & 0.3 & 0.0 & 1.0 \\
		$\delta m^\mathrm{sea}_{uds}\to0$               & 0.2 & 0.1 & 0.0 & 0.0 \\
		$\delta m^\mathrm{sea}_c\to0$               	& 0.3 & 0.1 & 0.0 & 0.0 \\
		$m_h \ne m_c$ (\eq{eq:Rn-4})         			& 0.1 & 0.1 & 0.0 & 0.0 \\
		Uncertainty in $w_0$, $w_0/a$                   & 0.2 & 0.0 & 0.1 & 0.4 \\
		$\alpha_0$ prior                                & 0.0 & 0.1 & 0.0 & 0.0 \\
		Uncertainty in $m_{\eta_s}$                     & 0.0 & 0.0 & 0.4 & 0.0 \\
		$m_h/m_c \to m_b/m_c$ 							& 0.0 & 0.0 & 0.0 & 0.4 \\
		$\delta m_{\eta_c}$: electromag., annih.        & 0.1 & 0.0 & 0.1 & 0.1 \\
		$\delta m_{\eta_b}$: electromag., annih.		& 0.0 & 0.0 & 0.0 & 0.1 \\
		\hline
		Total:  							& 0.64\% & 0.63\% & 0.55\% & 1.20\%
	\end{tabular}
	\end{ruledtabular}
\end{table}

\begin{figure}
	\includegraphics[scale=0.9]{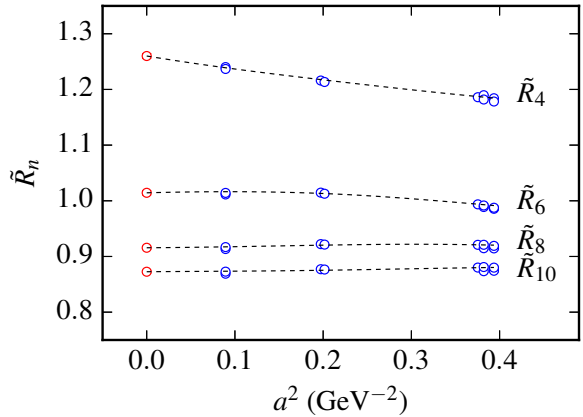}
	\caption{Lattice-spacing dependence of reduced moments $\tilde R_n$
	for $\eta_h$~masses within 5\% of $m_{\eta_c}$, and $n=4$, 6, 8, 10.
	The dashed lines show our fit,
	and the points at $a=0$ are the continuum extrapolations of the
	lattice data.}
	\label{fig:Rn-a2}
\end{figure}

\begin{figure}
	\includegraphics[scale=0.9]{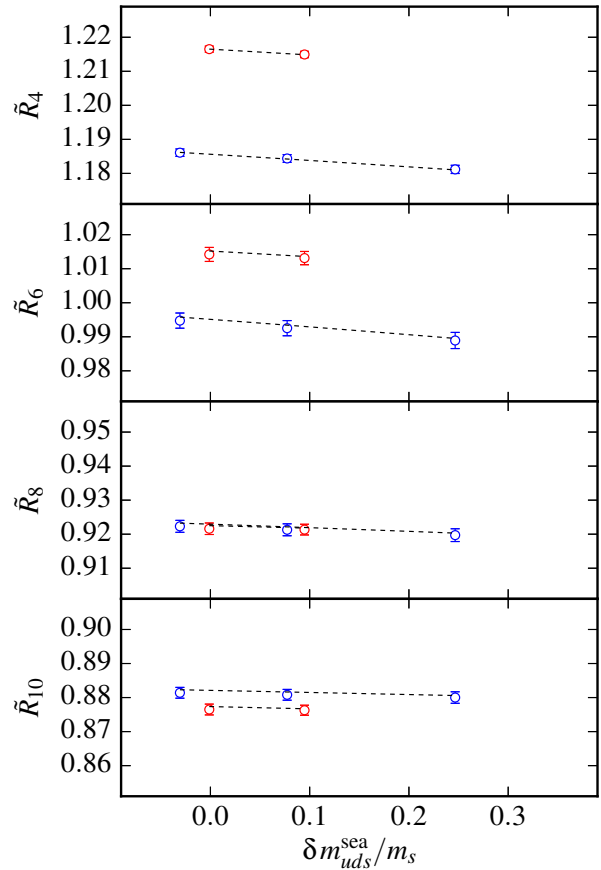}
	\caption{Light sea-quark mass dependence of reduced moments $\tilde R_n$
	for $m_h=m_c$, and $n=4$, 6, 8, 10. Results are shown for our
	two coarsest lattices: $a=0.12$\,fm (three points in blue) 
	and $a=0.09$\,fm (two points in red).
	The dashed lines show the corresponding results from our fit.
	Note that the slopes of the lines are independent of the lattice spacing,
	as expected.
	}
	\label{fig:Rn-msea}
\end{figure}

The dominant sources of error for our results are listed in 
Table~\ref{tab:nf4errors}.
The most important systematics are due to the
truncation of perturbation theory and our extrapolation to $a^2=0$. 
As in our previous analysis, the
$a^2$ extrapolations are not large, as is clear from
Figure~\ref{fig:mc-mh} and also Figure~\ref{fig:Rn-a2}.
Also the dependence of our results on the light sea-quark masses is 
quite small and independent of the 
lattice spacing, as illustrated by Figure~\ref{fig:Rn-msea}.

Our results change by~$\sigma/3$
if we fit only the $n=4$ and~6 moments, but the errors are 35\%~larger. 
Leaving out $n=4$, instead, leaves
the $c$~mass almost unchanged, but increases the error in the coupling
by~60\% (with the same central value).
We limit our analysis to heavy 
quark masses with $am_{0h}\le0.8$, as in our previous analysis.
Reducing that limit to~$0.7$,
for example, has no impact on the central values of results and 
increases our
errors only slightly (less than 10\%). 

\begin{figure}
	\includegraphics[scale=0.9]{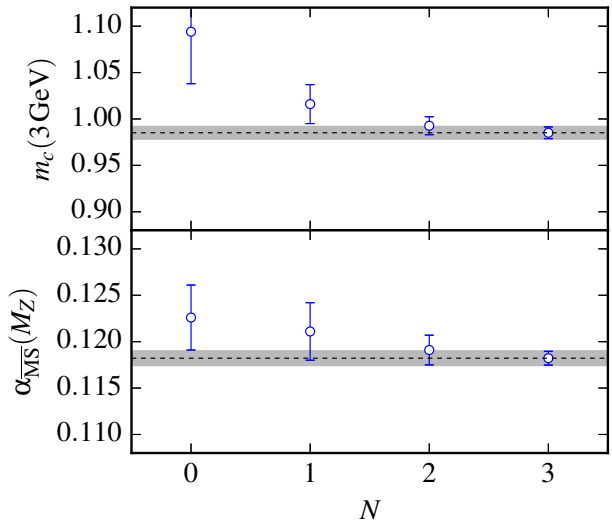}
	\caption{Results for the $\msb$ $c$~mass and coupling from 
	$n_f=4$ fits that treat perturbative coefficients 
	beyond order~$N$ as fit parameters, with priors specified by~\eq{eq:rnprior}.
	The gray bands and dashed lines indicate the means and standard deviations
	of our final results, which correspond to $N=3$.}
	\label{fig:errorbars}
\end{figure}

We tested the reliability of our error estimates for the
perturbation theory by refitting our data
using only a subset of the known perturbative coefficients. The results
are presented in Fig.~\ref{fig:errorbars}, which shows values 
for $m_c(3\,\mathrm{GeV})$
and $\almsb(M_Z)$ from fits 
that treat perturbative coefficients beyond
order~$N$ as fit parameters, with priors as in~\eq{eq:rnprior}.
Results from different orders agree with each other, 
providing evidence that our estimates
of truncation errors are reliable. This plot also shows the steady 
convergence of perturbation theory as additional orders are added.

As a further test of perturbation theory, we refit our
nonperturbative data treating the 
leading perturbative coefficients,~$\gamma_0$ and~$\beta_0$,
in the evolution equations for the mass~(\eq{eq:mh-evolution}) 
and coupling~(\eq{eq:almsb-evolution}) as fit parameters 
with priors of~$0\pm1$. The fit gives
\begin{equation}
	\gamma_0=0.292(19) \quad\quad \beta_0=0.675(54),
\end{equation}
in good agreement 
with the exact results of~$0.318$ and~$0.663$, respectively. So 
our nonperturbative results for the correlators show 
clear evidence for the evolution of $m_c(\mu)$ and~$\almsb(\mu)$
as $\mu=3m_h$ varies from~$3m_c$ to~$9m_c$.

\section{$m_c/m_s$ from $n_f=4$}

\begin{figure}
	\includegraphics[scale=0.9]{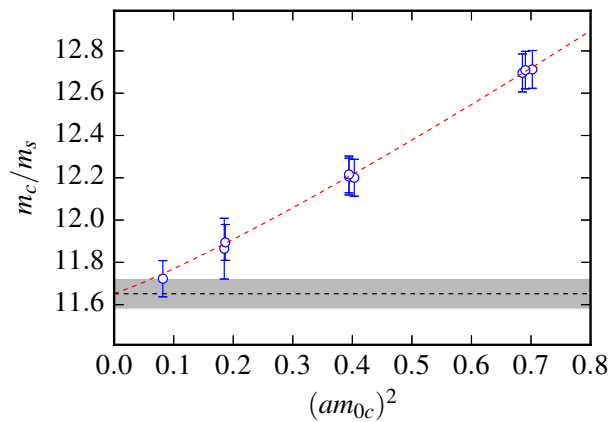} 
	\caption{The ratio of the $c$~and $s$~quark masses as a function of
	the squared lattice spacing (in units of the bare $c$ mass). 
	The data come from simulations at lattice spacings of 0.15, 0.12, 
	0.09 and~0.06\,fm, after tuning the $s$~and $c$~masses to 
	reproduce physical values for the~$\eta_s$ and $\eta_c$~masses
	on each ensemble.
	The errors for the data points are highly correlated, as they come 
	primarily from uncertainties in~$w_0$, $m_{\eta_s}$, and~$m_{\eta_c}$.
	The red dashed line shows our fit, which has a $\chi^2$~per degree of freedom
	of~0.21 for 9~degrees of freedom ($p$-value of 0.99). 
	The black dashed line and gray band show the mean value
	and standard deviation for our result extrapolated to zero lattice
	spacing. 
	}
	\label{fig:mc-ms}
\end{figure}

As discussed above (Section~\ref{sec:heavy-quark-correlators}),
we can use lattice QCD to extract ratios of $\msb$ quark masses completely
nonperturbatively~\cite{Davies:2009ih}, since ratios of quark masses are scheme 
and scale independent: for example,
\begin{equation}
	\left.
	\frac{m_{0c}}{m_{0s}}
	\right|_\mathrm{lat} 
	= \left.
	\frac{m_c(\mu,n_f)}{m_s(\mu,n_f)} 
	\right|_{\msb}
	+ \order((am_c)^2\alpha_s).
\end{equation}
While ratios of light-quark masses can be obtained from chiral 
perturbation theory, only lattice QCD can produce nonperturbative 
ratios involving heavy quarks. These ratios are very useful for checking 
mass determinations that rely upon perturbation theory, as illustrated
in~\cite{McNeile:2010ji}.
They also allow us to 
leverage precise values of 
light-quark masses from very accurately determined heavy-quark masses.

In~\cite{Davies:2009ih} we used nonperturbative simulations, with
$n_f=3$ sea quarks, to determine
the $s$~quark's mass from the $c$ quark's mass and the ratio $m_c/m_s$.
We repeat that analysis here, 
but now for $n_f=4$ sea quarks, using the tuned values
of the bare $s$ and $c$ masses for each of our lattice ensembles:
$am_{0s}^\mathrm{tuned}$ and 
$am_{0c}^\mathrm{tuned}$ in Table~\ref{tab:ensembles},
respectively. We expect
\begin{align}	
	\frac{am_{0c}^\mathrm{tuned}}{am_{0s}^\mathrm{tuned}}
	=&\, \frac{m_c}{m_s} 
	\left(
	1 
	+ h_m \frac{\delta m_{uds}^\mathrm{sea}}{m_s} 
	+ h_{a^2,m} \frac{\delta m_{uds}^\mathrm{sea}}{m_s} \left(
	\frac{m_c}{\pi/a}
	\right)^2 \right.
	\nonumber \\
	&\left.
	+ h_1 \alpha_s(\pi/a) \left(
	\frac{m_c}{\pi/a}
	\right)^2 
	+ \sum_{j=2}^{N_{a^2}} h_j \left(
	\frac{m_c}{\pi/a}
	\right)^{2j}
\right),
\end{align}	
where again we ignore $\delta m_c^\mathrm{sea}$ and $\delta m^2$ 
dependence since they are 
negligible. We fit the data from Table~\ref{tab:ensembles} using this 
formula with the following fit parameters and priors:
\begin{align}
	h_m &= 0 \pm 0.1, & h_{a^2,m} &= 0 \pm 0.1, \\
	h_1 &= 0 \pm 6, & h_j &= 0 \pm 2\quad (j>1).
\end{align}
The extrapolated value $m_c/m_s$ is also a fit parameter. 
We set $N_{a^2}=5$, but get identical results for any~$N_{a^2}\ge2$.

The result of this fit is presented in Fig.~\ref{fig:mc-ms}, which
shows the $a^2$ dependence of the lattice results. The sensitivity 
of our new results to~$a^2$
is about half what we saw in our previous analysis.
Our new fit is 
excellent and gives a final result for the mass ratio of:
\begin{equation}
	\frac{m_c(\mu,n_f)}{m_s(\mu,n_f)} = 11.652(65).
\end{equation}
The leading sources of error in this result are listed in 
Table~\ref{tab:nf4errors}. These are dominated by 
statistical errors and uncertainty in the $\eta_s$~mass.
Many other potential sources of error, such as uncertainties
in the lattice spacing, largely cancel in the ratio.

Note that the discussion in Appendix~\ref{app-msea} and 
\eq{eq:m-tilde}, in particular, imply that the leading 
effect of mistuned sea-quark
masses cancels in ratios of quark masses. This is substantiated
by our fit which makes parameter~$h_m$ negligibly small~($-0.0080(34)$).
Setting $h_m=0$ shifts our result for $m_c/m_s$ by only~$\sigma/7$.

Our result is a little more than a standard deviation lower than the recent result,
$11.747(19)\binom{+59}{-43}$, 
computed by the Fermilab/MILC collaboration (using many of 
the same configurations we use)~\cite{Bazavov:2014wgs}. 
Our analysis uses a different 
scheme for tuning the lattice spacing and quark masses, which leads to
the lack of sea-quark mass dependence in $m_c/m_s$ discussed just above. 
The absence of sea-mass dependence is apparent from Fig.~\ref{fig:mc-ms},
where the clusters of data points correspond to ensembles with the same
bare lattice coupling but different sea-quark masses. This figure 
can be compared with Fig.~6 in~\cite{Bazavov:2014wgs}, which shows much 
larger sea-mass dependence. Both approaches should agree
when extrapolated to zero lattice spacing and the physical 
sea-quark masses.

\section{$m_h/m_c$ from $m_{\eta_h}$}
\label{sec:mh/mc}
An analysis similar to that in the previous section allows us to relate
heavy-quark masses $m_h$ to the $h\overline h$~pseudoscalar mass 
$m_{\eta_h}$ with data from Table~\ref{tab:metah-Rn}. This can be
used, for example, to estimate the $b$~mass by extrapolating 
to $m_{\eta_b}$. 

\begin{figure}
	\includegraphics[scale=0.9]{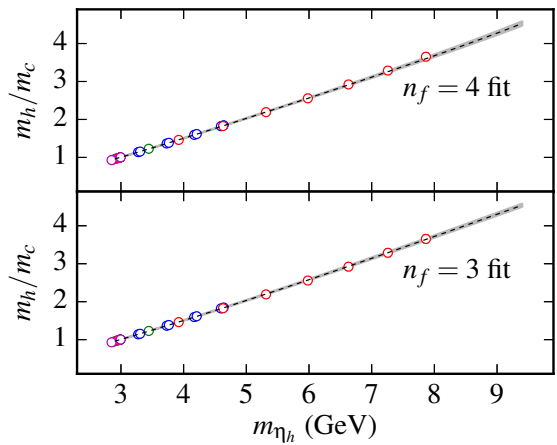} 
	\caption{The ratio of the $h$~and $c$~quark masses as a function 
	of the mass of $h\overline h$~pseudoscalar meson mass.
	The data come from simulations at lattice spacings of 0.15, 0.12, 
	0.09 and~0.06\,fm; the data points are colored magenta, blue, green, 
	and red, respectively. The gray band and dashed line in the top 
	panel show function~\eq{eq:mhmc-fitfcn} with the best fit parameters,
	extrapolated to zero lattice spacing and the correct sea-quark masses.
	The bottom panel compares the $n_f=4$~data with extrapolated results obtained
	in~\cite{McNeile:2010ji} from current-current correlators 
	in $n_f=3$~simulations.
	}
	\label{fig:mh-mc}
\end{figure}

Here we fit the lattice mass ratios $m_{0h}/m_{0c}^\mathrm{tuned}$ 
to the following function of $m_{\eta_h}$ from the simulation:
\begin{align}
\label{eq:mhmc-fitfcn}
	\frac{m_h}{m_c} &= \frac{m_{\eta_h}}{m_{\eta_c}}
	\sum_{n=0}^{N}
	f_n(m_{\eta_h}) \left(\frac{am_{\eta_h}}{4}\right)^{2n} \nonumber \\
	&+ f_\mathrm{sea}(\eta_h) \frac{m_{\eta_h}}{m_{\eta_c}}
	\frac{\delta m_{uds}^\mathrm{sea}}{m_s} 
	\left(\frac{am_{\eta_h}}{4}\right)^2
\end{align}
where $N=20$, although any $N>3$ gives the same result.
Here $f_n(m_{\eta_h})$ and $f_\mathrm{sea}(m_{\eta_h})$ are cubic splines
with knots at
\begin{equation}
	m_\mathrm{knots} = \knots\,\mathrm{GeV}.
\end{equation}
The maximum and minimum knots correspond to the maximum and minimum 
values of $m_{\eta_h}$, while the locations of the internal knots were 
obtained by treating those locations as fit parameters. 
Each $f$ is parameterized by
\begin{equation}
	f(m) = f_{0} + \delta f(m)
\end{equation}
and fit parameters
\begin{align}
	f_{0} &= 0 \pm 1 \nonumber \\
	\delta f(m) &= 0 \pm 0.15 & \quad m &\in m_\mathrm{knots} \nonumber\\
	\delta f^\prime(m) &= 
			0.15 \pm 0.15 &\quad m&=2.9\,\mathrm{GeV}.
\end{align}
We reduce the
priors for the leading $a^2$~errors by a factor of~$1/3$ since 
these errors are suppressed by~$\alpha_s$ in the HISQ~discretization.
The choice of priors for the spline parameters is motivated by 
results from~\cite{McNeile:2010ji} (see Figure~4 in that paper).

The fit is excellent with a $\chi^2$ per degree of freedom of~0.44 for
29~pieces of data: see the top panel in~Figure~\ref{fig:mh-mc}. 
Finite lattice spacing 
errors are much smaller for this quantity than for the moments, and it 
is again largely independent of mistunings in the sea-quark masses. 
Extrapolating to $m_{\eta_b}$ gives
\begin{equation}
	m_b/m_c = \mbmc
\end{equation}
which agrees with our $n_f=3$~result of~$4.51(4)$, but with larger
errors~\cite{McNeile:2010ji}. 
Our new $n_f=4$ data go down to lattice spacings of~$0.06$\,fm;
our earlier analysis also had results at~$0.045$\,fm.

The bottom panel of Figure~\ref{fig:mh-mc} compares our new
$n_f=4$~data with $n_f=3$~results
obtained from fits to the current-current correlators~\cite{McNeile:2010ji}. 
The agreement
is excellent, showing again that $n_f=3$ and $n_f=4$ are consistent
with each other.

\section{Conclusions and Outlook}

The initial extractions of quark masses from heavy-quark current-current
correlators  relied upon experimental data 
from $e\overline e$~annihilation~\cite{Chetyrkin:2009fv,Chetyrkin:2010ic}.
Our  analysis here, like the two that preceded
it~\cite{Allison:2008xk,McNeile:2010ji}, replaces experimental data with
nonperturbative results from tuned lattice simulations. 

Lattice simulations
offer several advantages over experiment  for this kind of
calculation~\cite{Lepage:2014fla}. 
For one thing,  simulations are easier to
instrument than experiments and much more flexible.  
Thus we can generate
lattice  ``data'' not just for vector-current correlators, but for any 
heavy-quark current or  density; 
we optimize our simulations by using the
pseudoscalar  density instead of 
the vector current. Experiment provides
results for  only two heavy-quark 
masses\,---\,$m_c$ and $m_b$\,---\,but we can
produce lattice data for a whole range of 
masses between $m_c$ and $m_b$.
This  means that $\almsb(\mu)$ varies 
continuously, by almost a factor of
two,  in our analysis since~$\mu\propto m_h$. 
Here we use this variation 
to estimate and bound uncalculated terms in perturbation theory, 
providing much  more reliable estimates of perturbative errors 
than the standard  procedure of replacing~$\mu$ by~$\mu/2$ and~$2\mu$. 
(Our analysis is essentially independent of~$\mu$.) 
Nonperturbative contributions are also strongly dependent upon~$m_h$, 
and therefore more readily bound if a range of 
masses is available; they
are negligible in our analysis.

In this paper,
we have redone our earlier $n_f=3$~analysis~\cite{McNeile:2010ji}
using simulations with
$n_f=4$ sea quarks: $u$, $d$, $s$ and~$c$. Our new results,
\begin{align}
	\label{eq:mc-result}
	m_c(3\,\mathrm{GeV}, n_f=4) &= \mcref\,\mathrm{GeV} \\
	\almsb(M_Z, n_f=5) &= \almz,	
\end{align}
agree well with our earlier results of $0.986(6)$\,GeV and $0.1183(7)$,
suggesting that contributions 
from $c$~quarks in the sea are reliably
estimated using perturbation theory (as expected). 
Our $c$~mass is about $1.8\sigma$ 
lower than the recent result from the ETMC collaboration, also
using $n_f=4$ simulations but 
with a different method~\cite{Carrasco:2014cwa}: 
they get $m_c(m_c)=1.348(42)$\,GeV, compared
with our $n_f=4$~result of~\mcmc\,GeV.

\begin{figure}
 	\includegraphics[scale=0.9]{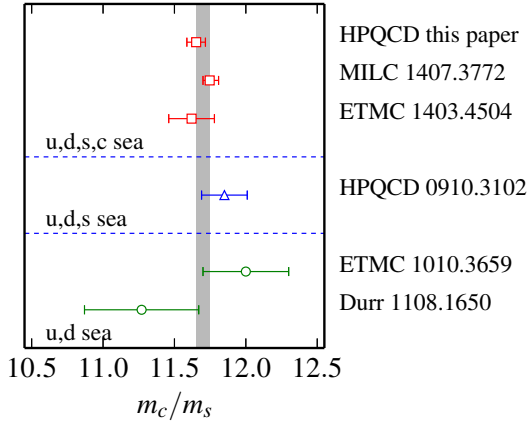}	
	 \caption{Lattice QCD determinations of the ratio of the $c$~and $s$ 
	 quarks' masses. The ratios come from this paper and
	references~\cite{Bazavov:2014wgs,Carrasco:2014cwa,Davies:2009ih,%
	Durr:2011ed,Blossier:2010cr}. 
	 The gray band is the weighted average
	 of the three $n_f=4$ results:~$11.700(46)$.
	 }
 	\label{fig:mc-ms-comparison}
 \end{figure}

We updated our earlier $n_f=3$ analysis~\cite{Davies:2009ih} 
of the ratio~$m_c/m_s$ of 
quark masses using our $n_f=4$~data. This is a relatively simple 
analysis of data from Table~\ref{tab:ensembles}. Our new value is:
\begin{equation}
	\frac{m_c(\mu,n_f)}{m_s(\mu,n_f)} = 11.652(65).	
\end{equation}
It agrees well with our previous result~$11.85(16)$, but is much more 
accurate. We compare our new result with others 
in Fig.~\ref{fig:mc-ms-comparison}.

We obtain a new estimate for the $s$~mass by combining our 
new result for $m_c/m_s$ with our new estimate of the 
$c$~mass (\eq{eq:mc-result}, converted from~$n_f=4$):
\begin{equation}
	m_s(\mu,n_f=3) = 
	\begin{cases}
		\mslow\,\mathrm{MeV} & \mu=2\,\mathrm{GeV} \\
		\mshigh\,\mathrm{MeV} & \mu=3\,\mathrm{GeV}.   
	\end{cases}
\end{equation}
Values for $m_s(\mu,n_f=4)$ are smaller by about~0.2\,MeV.
Our new result agrees with our previous analysis and also
with other recent $n_f=3$~or~4 analyses: 
\begin{align}
	m_s(2\,\mathrm{GeV}) &=
	\begin{cases}
		92.4(1.5)\,\mathrm{MeV} & \mbox{HPQCD~\cite{Davies:2009ih}},\\
		99.6(4.1)\,\mathrm{MeV} & \mbox{ETMC~\cite{Carrasco:2014cwa},} \\
		95.5(1.9)\,\mathrm{MeV} & 
		\mbox{Durr~\emph{et al}~\cite{Durr:2010vn},}
		\\
	\end{cases} \nonumber \\
	m_s(3\,\mathrm{GeV}) &= 83.5(2.0)\,\mathrm{MeV} \quad
	\mbox{RBC/UKQCD~\cite{Arthur:2012opa}}.
\end{align}

Finally, we have also updated our previous ($n_f=3$) 
nonperturbative analysis of $m_b/m_c$ using our new $n_f=4$~data. We
obtain:
\begin{equation}
	\frac{m_b(\mu,n_f)}{m_c(\mu,n_f)} = \mbmc,
\end{equation}
which agrees with our previous result of 4.51(4)~\cite{McNeile:2010ji}.
Combining this result with our new value for $m_c$ (\eq{eq:mc-result}) gives
\begin{equation}
	m_b(m_b,n_f=5) = \mbmb.
\end{equation}
This again agrees with our earlier result of~4.164(23)\,GeV, but with larger
errors. We can also multiply our results for $m_b/m_c$ and $m_c/m_s$
to obtain
\begin{equation}
		\frac{m_b(\mu,n_f)}{m_s(\mu,n_f)} = \mbms.
\end{equation}
This is almost four standard deviations (but only 4\%) away from the 
result predicted by the Georgi-Jarlskog 
relationship~\cite{Georgi:1979df}
for certain classes of grand unified theory: the
Georgi-Jarlskog relationship says that
$m_b/m_s$ should equal $3m_\tau/m_\mu=50.45$.

The prospects for improving our results over the next decade are good. 
Detailed meta-simulations, described in~\cite{Lepage:2014fla}, indicate that 
errors from our analysis can be pushed below 0.25\% 
by a combination of higher-order 
perturbation theory, and, especially, smaller lattice spacings (0.045,
0.03 and~0.023\,fm)\,---\,both improvements that are quite 
feasible over a decade~\cite{Lepage:2014fla}. 
There are also many other promising approaches within lattice QCD.
Several exist already for extracting the QCD coupling: see, for example,
\cite{Davies:2008sw,Shintani:2010ph,Bazavov:2012ka,%
Jansen:2011vv,Fritzsch:2012wq,Blossier:2013ioa}.
One can also use simulations of other renormalized quantities, 
such as the $m_h\overline\psi_h\gamma_5\psi$ vertex function, to compute 
quark masses~\cite{Martinelli:1994ty}. 

Small lattice spacings are particularly important for the $b$~mass, 
because lattice spacing errors are typically of order~$(am_b)^2$. 
One approach is to use highly-improved relativistic actions for the 
$b$~quarks, like the HISQ action used here. As shown in~\cite{Follana:2006rc},
all but one of the $\order(a,a^2)$ operators that arise in the Symanzik improvement
of a quark action are suppressed by extra factors of the heavy-quark
velocity: factors of $(v/c)^2$ for mesons made of heavy quarks, and
$v/c$ for mesons made of a combination of heavy and light quarks.
The one operator that does not have extra suppression is
$\sum_\mu\overline\psi \gamma^\mu (D^\mu)^3\psi$, 
which violates Lorentz invariance
and so is easily tuned nonperturbatively using the meson dispersion relation.
This is the strategy adopted in the HISQ discretization we use here. The extra
factors of $v/c$ suppress $(am_b)^2$ errors by an extra order of magnitude,
beyond the suppression, by a power of~$\alpha_s$, coming from tree-level 
corrections for $a^2$ errors in HISQ.

$(am_b)^2$ errors can be avoided completely by using effective field theories
like NRQCD~\cite{Lee:2013mla} or the Fermilab formalism~\cite{ElKhadra:1996mp}
for $b$~dynamics. Such approaches should be sufficiently accurate 
provided  they are corrected to sufficiently high order in~$(v_b/c)^2$.
Our recent NRQCD analysis of~$m_b$, using current-current correlators,
is encouraging~\cite{Colquhoun:2014ica}.

Overall the prospects are excellent for continued improvement.

\begin{acknowledgments}
We are grateful to the MILC collaboration 
for the use of their 
gauge configurations and code. 
We thank S.~King and D.~Toussaint for useful conversations.
Our calculations were done on the Darwin Supercomputer 
as part of STFC's DiRAC facility jointly
funded by STFC, BIS 
and the Universities of Cambridge and Glasgow. 
This work was funded by STFC, the Royal Society, the Wolfson Foundation
and the National Science Foundation.
\end{acknowledgments}

\appendix 

\section{Sea-Quark Mass Dependence}
\label{app-msea}

In this appendix we discuss the dependence of the $\msb$ coupling and 
heavy-quark masses on the sea-quark masses. 
We vary the $u/d$ sea-quark mass in our simulations 
to help us assess systematic errors associated with tuning 
that mass. In addition, the precision with which the $s$ and $c$ sea-quark masses 
have been tuned varies by several percent over the various ensembles we use. 
These detunings shift the $\msb$~coupling and masses. 
We need to understand how they 
are shifted in order to extract results for~$\almsb$ and~$m_h$ with
physical sea-quark masses. 

It is essential when discussing detuned sea-quark masses to be specific about
what is held fixed as the quark masses are shifted from their physical values.
An obvious choice is to fix both the lattice spacing $a$ and 
the bare coupling $\alpha_\mathrm{lat}$ in the lattice lagrangian, while varying
the quark masses. We find it more convenient, however, to 
explore a slightly different manifold in theory space 
by fixing $\alpha_\mathrm{lat}$
and the value of the Wilson-flow parameter~$w_0$. 

Lattice simulations
are done for particular values of the bare coupling constant (and
bare quark masses), but with all dimensional quantities expressed 
in units of the lattice spacing (\emph{lattice units}).
This removes explicit dependence on the lattice spacing from the simulation,
so we can run the simulation without knowing the lattice spacing. To 
extract physics, however, we must determine the lattice spacing
(from the simulation) and convert all simulation results from lattice units
to physical units.
In our simulations, we calculate the lattice spacing by measuring the value  
of~$a/w_0$ in the simulation, and multiplying it by the known value 
of $w_0$~for physical sea-quark masses (that is, 0.1715(9)\,fm). As a result
the lattice spacing becomes (weakly) dependent upon the sea-quark masses
since $w_0$ is affected by sea quarks.

This procedure is convenient because the lattice spacing for a given ensemble
is determined using information from only that ensemble, thereby decoupling
the analyses of different ensembles to a considerable extent. As we discuss 
below there is an added benefit when vacuum polarization from~$c$
(or heavier) quarks is included in the simulation, as we do here: 
heavy quarks automatically 
decouple from low-energy physics (like 
$w_0$~\cite{*[{Perturbation theory 
suggests that the important scales 
in~$w_0$ are of order
$Q_{w_0} = 1/\sqrt{8w^2_0}\approx400$\,MeV. 
See }][] Luscher:2010iy}). 
With our procedure, 
physical quantities that probe energy scales smaller than $2m_c$\,---\,that is,
almost everything studied with lattice QCD today\,---\,are essentially 
independent of~$m_c$, which means that they are completely 
unaffected by tuning errors in~$m_c$. 
This would not be the case if we 
fixed the lattice spacing instead of~$w_0$, since it is small variations in 
the lattice spacing that correct for mistuning in~$m_c$.

It is also very convenient that we set the lattice spacing using a 
flavor singlet quantity.
Because~$w_0$ is a flavor singlet, the leading 
sea-mass dependence induced in the lattice 
spacing is analytic (linear) in the quark mass and small; 
in particular, there are no chiral logarithms~\cite{Bar:2013ora}. 
One consequence is that
leading-order chiral perturbation theory for physical quantities 
($f_\pi$, $f_{D_s}$\ldots) is unchanged from standard treatments except
for shifts (that are easily accommodated) in the coefficients 
of certain analytic terms.

In this appendix we show how the
$\msb$~coupling and heavy-quark mass depend upon the sea-quark masses in our
simulations. This dependence implies sea-quark mass 
dependence in the lattice spacing and the heavy quark's bare mass, 
which we then use to determine some of the parameters involved. 
Finally we review heavy-quark decoupling, and 
estimate the parameters for $c$-mass dependence 
using first-order perturbation theory.

\subsection{Tuning Bare Quark Masses}
\label{sec:tuning-masses}
We define tuned values for the bare~$c$ and $s$~masses on each ensemble
by adjusting those masses to give physical values in simulations
for the $\eta_c$ and $\eta_s$~masses. The tuned values are listed in
Table~\ref{tab:ensembles}.

The current experimental value for the $\eta_c$~mass
is~2.9836(7)\,GeV~\cite{Agashe:2014kda}. In our analysis, we remove 
electromagnetic corrections from this value, 
and adjust its error to account for $c\overline c$~annihilation,
since neither effect is in our 
simulations~\cite{fdsupdate,gregory}. We use:
\begin{equation}
	m_{\eta_c}^\mathrm{phys} = 2.9863(27)\,\mathrm{GeV}.
\end{equation}
We compute the tuned $c$~mass $m_{0c}^\mathrm{tuned}$ 
by linear interpolation 
using $\eta_h$~masses from the simulation (Table~\ref{tab:metah-Rn}) 
for heavy-quark masses~$m_{0h}$ in the vicinity of~$m_{0c}$.
In a few cases we have results for only a single value of~$m_{0h}$;
then we compute the tuned $c$~mass using estimates of 
$dm_{\eta_c}/dm_{0c}$ from other ensembles with (almost) the same lattice 
spacing. 

Note that the uncertainty in $m_{0c}^\mathrm{tuned}$ is usually 
\emph{smaller} than that in $am_{0c}^\mathrm{tuned}$. This is a peculiar
feature of heavy-quark masses in lattice simulations 
(see, for example,~\cite{Davies:1994pz}). It follows from the formula
for the linear interpolation that defines the tuned mass in terms of
a nearby mass:
\begin{equation}
\label{eq:mc-tuned}
	m_{0c}^\mathrm{tuned} =  (am_{0c}) a^{-1} + \frac{dm_{0c}}{dm_{\eta_c}}
	\left(m_{\eta_c}^\mathrm{phys} -  (am_{\eta_c}) a^{-1}\right)
\end{equation}
where $am_{\eta_c}$ is the simulation result for the $\eta_c$ mass (in lattice units)
when the $c$~quark has mass $am_{0c}$. Here $dm_{0c}/dm_{\eta_c}$ is obtained
from simulation results for a few nearby $c$~masses. The uncertainty in
$a^{-1}$ is usually larger than the uncertainties in the other 
lattice quantities, but here $a^{-1}$ is multiplied by
\begin{equation}
	(am_{0c}) - (am_{\eta_c}) \frac{dm_{0c}}{dm_{\eta_c}}
\end{equation}
which would vanish if $m_{\eta_c} = 2m_{0c}$. This cancellation is only
partial for real masses, but it doesn't occur at all if~\eq{eq:mc-tuned}
is multiplied on both sides by~$a$ to give a formula for~$am_{0c}^\mathrm{tuned}$. 
As a result, fractional errors are roughly $3\times$~smaller 
for~$m_{0c}^\mathrm{tuned}$.

The $\eta_s$ is an 
$s\overline s$~pseudoscalar particle where the valence quarks are (artificially) 
not allowed to annihilate; its physical mass is determined 
in lattice simulations from the masses of the pion 
and kaon~\cite{Dowdall:2013rya}:
\begin{equation}
		m_{\eta_s}^\mathrm{phys} = 0.6885(22)\,\mathrm{GeV}
\end{equation}
This mass is defined for use in lattice simulations and needs no further
corrections for electromagnetism.
We tune the 
$s$~mass by simulating with a nearby bare mass~$m_{0s}$ 
to obtain the
corresponding $\eta_s$~mass, and then extracting the tuned mass using:
\begin{equation}
	\label{eq:tuned-ms}
	m_{0s}^\mathrm{tuned} = m_{0s} \left(
	\frac{m_{\eta_s}^\mathrm{phys}}{m_{\eta_s}}
	\right)^2.
\end{equation}
Our $\eta_s$~data are presented in 
Table~\ref{tab:etas}, which shows that the
tuned mass is quite insensitive to small variations in~$m_{0s}$.
We do not have $\eta_s$ results for ensemble~7; 
there the tuned $s$ mass is based
on an interpolation between results 
from ensemble~8 and another ensemble
that has similar parameters but with $am_{0\ell}=0.0074$. 

Table~\ref{tab:ensembles} shows that $m_{0c}^\mathrm{tuned}$ is 
more accurate than $m_{0s}^\mathrm{tuned}$. This is because the 
uncertainties in the value of the lattice spacing have a smaller 
impact on the $c$~mass because the cancellation described above 
only happens for heavy quarks (where $m_{\eta_h}\approx 2 m_{0h}$).

\begin{table}
	\caption{Simulation results for the $\eta_s$ mass $am_{\eta_s}$ 
	corresponding to different values of 
	the bare $s$~mass $am_{0s}$ and different gluon ensembles. 
	The ensembles are described in Table~\ref{tab:ensembles},
	although we use many more configurations for our~$\eta_s$
	analysis than are indicated there. Estimates for the tuned bare $s$~mass
	(\eq{eq:tuned-ms}) are also given.
	}
	\label{tab:etas}
	\begin{ruledtabular}
		\begin{tabular}{cclc}
		ensemble & $am_{0s}$ 
		& $\quad am_{\eta_s}$ & $am_{0s}^\mathrm{tuned}$
		\\ \hline
 1&  0.0705  &     0.54024(15)  &       0.0700(9)\\
  &  0.0688  &     0.53350(17)  &       0.0700(9)\\
  &  0.0641  &     0.51511(16)  &       0.0700(9)\\
 2&  0.0679  &      0.52798(9)  &       0.0686(8)\\
  &  0.0636  &      0.51080(9)  &       0.0687(8)\\
 3&  0.0678  &      0.52680(8)  &       0.0677(8)\\
\hline 
 4&  0.0541  &     0.43138(12)  &       0.0545(7)\\
  &  0.0522  &     0.42358(11)  &       0.0545(7)\\
 5&  0.0533  &      0.42637(6)  &       0.0533(7)\\
  &  0.0507  &     0.41572(14)  &       0.0534(7)\\
  &  0.0505  &      0.41474(8)  &       0.0534(7)\\
 6&  0.0527  &      0.42310(3)  &       0.0527(6)\\
  &  0.0507  &      0.41478(4)  &       0.0527(6)\\
\hline 
 8&  0.0360  &      0.30480(4)  &       0.0364(4)\\
\hline 
 9&  0.0231  &      0.20549(8)  &       0.0234(3)\\
		\end{tabular}
	\end{ruledtabular}
\end{table}

We set the $u$~and $d$~masses equal to their average,
\begin{equation}
	m_\ell \equiv \frac{m_u + m_d}{2},
\end{equation}
and set $m_\ell$~equal to the tuned $s$~mass (above) divided by
the physical value of the quark mass ratio~\cite{Bazavov:2014wgs}  
\begin{equation}
	\frac{m_s}{m_\ell} = 27.35(11).
\end{equation}

\subsection{$\almsb(\mu,\delta m^\mathrm{sea})$ and $a(\delta m^\mathrm{sea})$}
The beta function in the $\msb$~scheme is, by definition, independent of 
sea-quark masses. Thus the coupling's evolution is unchanged by 
detuned sea-quark masses\,---
\begin{equation}
	\label{eq:al-evol}
	\frac{d\almsb(\mu,\delta m^\mathrm{sea})}{d\log\mu^2} = 
	\beta(\almsb(\mu,\delta m^\mathrm{sea})) 
\end{equation}
---\,but mass dependence enters through the low-energy starting 
point for that evolution implied by the scale-setting procedure used
in the lattice simulation. Such mass dependence can enter only through 
an overall renormalization of the scale parameter~$\mu$:
\begin{equation}
	\label{eq:alpha-tilde}
	\alpha_{\overline{\mathrm{MS}}}(\mu, \delta m^\mathrm{sea}) 
	= \almsb(\xi_\alpha\mu) 
\end{equation}
where 
\begin{align}
	\almsb(\mu) &\equiv \almsb(\mu,\delta m^\mathrm{sea}=0)
\end{align}
is the $\msb$~coupling for physical sea-quark masses. The scale 
factor,
\begin{align}
	\label{eq:xi-alpha}
	\xi_\alpha \equiv  1 &
	+ g_\alpha\frac{\delta m_{uds}^\mathrm{sea}}{m_s}
	+ g_{a^2,\alpha} 
	\frac{\delta m_{uds}^\mathrm{sea}}{m_s} 
	\left(\frac{m_c}{\pi/a}\right)^2 
	\nonumber \\
	& + g_{c,\alpha} \frac{\delta m_{c}^\mathrm{sea}}{m_c} + \order(\delta m^2),
\end{align}
depends upon the differences between the masses $m_q$ used in the simulation
and the tuned values of those masses $m_q^\mathrm{tuned}$ 
(Table~\ref{tab:ensembles} and Sec.~\ref{sec:tuning-masses}):
\begin{align}
	\delta m_{uds}^\mathrm{sea} &\equiv \sum_{q=u,d,s} 
	\left(m_q - m_q^\mathrm{tuned} \right)\\
	\delta m_c^\mathrm{sea} &\equiv m_c - m_c^\mathrm{tuned}.
\end{align}
Function $\almsb(\xi_\alpha \mu)$ 
satisfies the standard evolution equation
(Eq.~(\ref{eq:al-evol})) because $\xi_\alpha$ is independent of~$\mu$.

We work to first order in $\delta m^\mathrm{sea}$ because higher-order 
terms are negligible in our simulations. As suggested above,
he leading-order dependence is
particularly simple because we use iso-singlet mesons ($\eta_c$ and $\eta_s$)
to set the $c$~and $s$~masses; in particular, there are no chiral logarithms
of the $u/d$~mass in leading order.

We expect coefficients $g_\alpha$ and $g_{a^2,\alpha}$ in $\xi_\alpha$
to be of order~$1/10$
since corrections linear in light-quark masses must be due to chiral symmetry
breaking and so should be of order $\delta m^{sea}/\Lambda$
where~$\Lambda\approx 10 m_s$. As we discuss below, $g_{c,\alpha}$
can be estimated from perturbation theory and is again of order~$1/10$.
We treat these coefficients as fit parameters in our analysis, with 
priors:
\begin{equation}
	\label{eq:g-alpha-prior}
	g_\alpha = 0 \pm 0.1, \quad g_{a^2,\alpha} = 0 \pm 0.1, \quad
	g_{c,\alpha} = 0 \pm 0.1.
\end{equation}

The rescaling factor~$\xi_\alpha$ is closely related to the dependence of 
the lattice spacing on the sea-quark masses used in the simulation. 
The lattice spacing is primarily a function of the bare 
coupling $\alpha_\mathrm{lat}$ used in the lattice action, but it also 
varies with the sea-quark masses, in our scheme, 
when the bare coupling is held constant.
As discussed above, this is because of sea-mass dependence 
in the quantity used to define the lattice spacing, $a/w_0$ in our case. 
The relationship with $\xi_\alpha$ 
can be understood by examining the $\msb$ coupling at scale~$\mu=\pi/a$.
There it is related to the bare coupling by a perturbative expansion,
\begin{align}
 	\almsb(\pi/a, \delta m^\mathrm{sea}) &= \almsb(\xi_\alpha \pi/a) 
 	\nonumber \\
 	&= \alpha_\mathrm{lat} + \sum_{n=2}^\infty c_n^\msb\,\alpha_\mathrm{lat}^n,
\end{align} 
that is mass-independent up to corrections 
of~$\order((am_c)^2\alpha_s)$, which are 
negligible in our analysis. This formula implies that $\almsb(\xi_\alpha \pi/a)$
is constant if $\alpha_\mathrm{lat}$ is, and therefore that $\xi_\alpha / a$ 
must be constant as well. Consequently the lattice spacing must vary 
with $\delta m^\mathrm{sea}$ like
\begin{align}
	\label{eq:a-msea}
	a(\delta m^\mathrm{sea})
	\approx \xi_\alpha\, a_\mathrm{phys}
\end{align}
if the bare coupling is held constant, where $a_\mathrm{phys}$~is 
the lattice spacing when the sea-quark masses are tuned to their 
physical values\,---\,that is, 
$a_\mathrm{phys} \equiv a(\delta m^\mathrm{sea}=0)$.

We use this variation in the lattice spacing 
to read off the parameters in~$\xi_\alpha$.
Our simulation results fall into four groups of gluon ensembles, 
with lattice spacings around 0.15\,fm, 0.12\,fm, 0.09\,fm and 0.06\,fm.
Each group corresponds to a single value of the 
bare lattice coupling $\alpha_\mathrm{lat}$,
and several different values of light sea-quark mass. Within a single group, then,
the values we obtain for $a/w_0$ from our simulations should vary as
\begin{equation}
	(a/w_0)_\mathrm{sim} = 
	\xi_\alpha \times (a/w_0)_\mathrm{phys},
\end{equation}
where the parameters
$g_{\alpha}$, $g_{a^2,\alpha}$ and $g_{c,\alpha}$ in $\xi_\alpha$ 
(\eq{eq:xi-alpha}) are the same for 
all four groups of data.

\begin{figure}
   \includegraphics[scale=0.9]{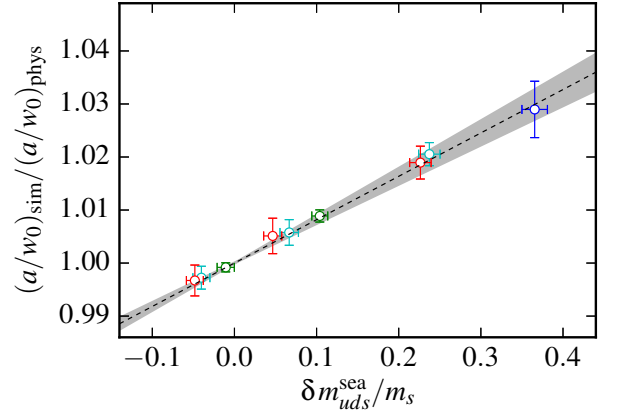}
   \caption{The ratio of the simulation lattice spacing with detuned sea-quark 
   masses to the lattice spacing with physical sea-quark masses as a function
   of the light-quark mass detuning (in units of the $s$~quark mass). 
   Results are shown for four different sets of
   data, each corresponding to a different bare lattice coupling. 
   The approximate lattice spacings 
   for these sets are: 0.15\,fm (red points),
   0.12\,fm (cyan), 0.09\,fm (green), and 0.06\,fm (blue). The dashed line
   and gray band show the mean and standard deviation of our best fit to these
   data. The fit has a~$\chi^2$ per degree of freedom of~0.23 for 9~degrees
   of freedom ($p$-value of 0.99).}
   \label{fig:a-w0}
\end{figure}

We fit our simulation results for $a/w_0$, simultaneously for all four groups,
as functions of $g_{\alpha}$, $g_{a^2,\alpha}$ and $g_{c,\alpha}$. 
We also treat the value of $(a/w_0)_\mathrm{phys}$ for each group as
a fit parameter.
The resulting fit is shown in Fig.~\ref{fig:a-w0} where we plot 
\begin{equation}
\frac{(a/w_0)_\mathrm{sim}}{(a/w_0)_\mathrm{phys}} \nonumber
\end{equation} 
versus $\delta m^\mathrm{sea}_{uds}/m_s$. 

The fit is excellent, and
shows that $g_\alpha=0.082(8)$. Our fit is not very sensitive
to $g_{a^2,\alpha}$ and $g_{c,\alpha}$\,---\,their impact on~$\xi_\alpha$ 
is too small\,---\,and gives results for these that are essentially
the same as the prior values.

\subsection{$m_h(\mu,\delta m^\mathrm{sea})$ and $m_{0c}(\delta m^\mathrm{sea})$}
The evolution equations for the heavy quark's $\msb$ mass are 
unchanged by sea-mass detunings: 
\begin{equation}
	\label{eq:m-evol}
	\frac{d\log(m_h(\mu,\delta m^\mathrm{sea}))}{d\log\mu^2} = 
	\gamma_m(\almsb(\mu,\delta m^\mathrm{sea}))
\end{equation}
Consequently any sea-mass dependence must enter through rescalings:
\begin{equation}
	\label{eq:m-tilde}
		m_h(\mu, \delta m^\mathrm{sea}) = \xi_m m_h(\xi_\alpha\mu)
\end{equation}
where $\xi_\alpha$ is defined above (Eq.~(\ref{eq:xi-alpha})), $\xi_m$~is
independent of~$\mu$, and
\begin{equation}
	m_h(\mu) \equiv m_h(\mu, \delta m^\mathrm{sea}=0)
\end{equation}
is the $\msb$~mass for physical sea-quark masses. We parameterize 
$\xi_m$ similarly to~$\xi_\alpha$ but allowing for the coefficients
to depend upon the heavy-quark mass:
\begin{align}
	\xi_m = 1 
	&+ \frac{g_m}{(m_{\eta_h}/m_{\eta_c})^\zeta} 
	\frac{\delta m_{uds}^\mathrm{sea}}{m_s}
	\nonumber \\
	&+ \frac{g_{a^2,m}}{(m_{\eta_h}/m_{\eta_c})^\zeta} 
		\frac{\delta m_{uds}^\mathrm{sea}}{m_s}
		\left(\frac{m_c}{\pi/a}\right)^2 
	+ \cdots
	\label{eq:xi-m}
\end{align}
Again we expect $g_m$ and $g_{a^2,m}$ to be of order~$1/10$, and we treat
them as fit parameters with priors:
\begin{equation}
	g_m = 0 \pm 0.1, \quad g_{a^2,m} = 0 \pm 0.1.
\end{equation}
We parameterize the dependence on heavy-quark mass with the factors
$(m_{\eta_h}/m_{\eta_c})^\zeta$ where $\zeta$ is a fit parameter
with prior:
\begin{equation}
	\zeta = 0 \pm 1.
\end{equation}

The sea-mass dependence in $\xi_m$ comes from the quantity used
to tune the heavy-quark mass in simulations. We tune these masses 
to give the correct physical mass for $\eta_h$\,---\,that is, the
mass obtained when the sea-quark masses are tuned to their physical
values and the lattice spacing is set to zero. This means that any 
sea-mass dependence in $m_{\eta_h}$ is pushed into the rescaling
factor~$\xi_m$ in \eq{eq:m-tilde}. The physical size of $\eta_h$~mesons
decreases as $m_{\eta_h}$ increases, and this decreases the 
coupling with light sea-quarks. Thus we expect $\zeta>0$ in~\eq{eq:xi-m};
our fit finds~$\zeta=0.3(1)$.

In principle, $\xi_m$ should depend upon~$\delta m_c^\mathrm{sea}$, 
as well as~$\delta m_{uds}^\mathrm{sea}$. Perturbation theory, however,
indicates that this dependence is negligible in our simulations. Thus
we have omitted such terms from~$\xi_m$.
We have verified that they are negligible by comparing fits that 
include $\delta m_c^\mathrm{sea}$ terms with 
the fit without them.

The rescaling factor $\xi_m$ is closely related to the sea-mass dependence
of the heavy quark's bare mass, in much the same way $\xi_\alpha$ is 
related to the lattice spacing. The bare mass $m_{0h}$ is proportional 
to the $\msb$~mass evaluated at $\mu=\pi/a$:
\begin{align}
	m_{0h} &\propto m_h(\pi/a,\delta m^\mathrm{sea}) \nonumber \\
	&\propto \xi_m m_h(\xi_\alpha \pi/a).
\end{align}
Since $\xi_\alpha/a$ is sea-mass independent, we see that $m_{h0}$
is proportional to $\xi_m$,
\begin{equation}
	m_{0h}(\delta m^\mathrm{sea}) = \xi_m m_{0h}^\mathrm{phys},
\end{equation}
when the sea-quark masses are varied while holding the bare coupling fixed.

This variation can be used to determine the parameters in~$\xi_m$, again
in analogy to the previous section. 
As discussed in the previous
section, our ensembles fall into four groups each corresponding to 
a different value of the bare coupling constant~$\alpha_\mathrm{lat}$. 
The masses $am_{0c}^\mathrm{tuned}$
for each ensemble
in Table~\ref{tab:ensembles} are tuned to give the physical $\eta_c$~mass
for that ensemble. Therefore, within each group of ensembles, we expect
\begin{equation}
		am_{0c}^\mathrm{tuned} = \xi_\alpha \xi_m \times (am_{0c})_\mathrm{phys}
\end{equation}
where $(am_{0c})_\mathrm{phys}$ is the value for properly tuned sea-quark
masses. 

We fit our simulation results for $am_{0c}^\mathrm{tuned}$ 
as functions of $g_m$, $g_{a^2,m}$,
$g_\alpha$, $g_{a^2,\alpha}$, and $g_{c,\alpha}$. We use best-fit values
from the fit in the previous section as priors for the last three of these
fit parameters. The values of $(am_{0c})_\mathrm{phys}$ for the different groups
of ensembles are also fit parameters. 

\begin{figure}
   \includegraphics[scale=0.9]{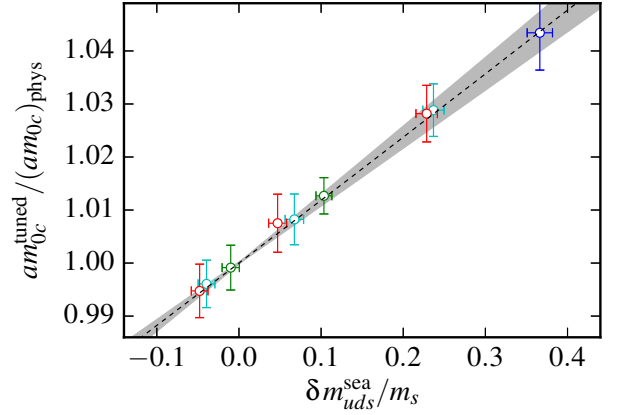}
   \caption{The ratio of the bare $c$~mass in lattice units used in the
   simulations to the bare mass with physical sea-quark masses as a function
   of the light-quark mass detuning (in units of the $s$~quark mass). 
   Results are shown for four different sets of
   data, each corresponding to a different bare lattice coupling. 
   The approximate lattice spacings for these sets are: 0.15\,fm (red points),
   0.12\,fm (cyan), 0.09\,fm (green), and 0.06\,fm (blue). The dashed line
   and gray band show the mean and standard deviation of our best fit to these
   data. The fit has a~$\chi^2$ per degree of freedom of~0.15 for 9~degrees
   of freedom ($p$-value of 1.0).}
   \label{fig:amc}
\end{figure}

The resulting fit is shown in Fig.~\ref{fig:amc},
where we plot $am_{0c}^\mathrm{tuned} / (am_{0c})_\mathrm{phys}$ as a 
function of $\delta m_{uds}^\mathrm{sea}/m_s$. The fit is excellent 
and shows that $g_m=0.035(5)$, while $g_{a^2,m}$ is 
essentially unchanged from its prior
value (because our data are not sufficiently accurate).

\subsection{$c$ Quarks and Decoupling}
Heavy quarks decouple from low-energy
physics, and therefore variations in $\delta m_c^\mathrm{sea}$ should have
no impact on physics (like $w_0$) that probes momentum scales 
smaller than~$m_c$. We can, however, introduce (apparent) violations of 
the decoupling theorem through the scheme used to set the lattice 
spacing. In particular, decoupling is violated by any scheme that holds 
the lattice spacing fixed (together with the bare 
coupling~$\alpha_\mathrm{lat}$) as $\delta m_c^\mathrm{sea}$ is varied.
On the contrary, decoupling is preserved by schemes that hold 
a low-energy~($<2m_c$) quantity like~$w_0$ fixed, 
instead of the lattice spacing~\footnote{Dimensionless ratios of low-energy 
quantities are independent of the lattice-spacing scheme, and must be 
independent of~$m_c$ by the decoupling theorem. 
This means that a scheme that makes any one low-energy quantity\,---\,for example,
$w_0$\,---\,independent of~$m_c$ makes all other low-energy
quantities independent of~$m_c$
as well, thereby preserving decoupling.}.

The difference between these schemes arises because the running
of the QCD coupling is modified in a detuned theory for 
scales between $m_c^\mathrm{sea}$ 
and $m_c^\mathrm{sea}+\delta m_c^\mathrm{sea}$, resulting in a mismatch
between low and high energy values of the coupling. 
Physics below~$m_c$ is determined by the $n_f=3$ coupling constant, which, 
by decoupling, should be independent of~$\delta m_c^\mathrm{sea}$.

To see how this works, we examine lowest-order perturbation theory
where
\begin{equation}
	\alpha_s^{(n_f)}(\mu) = \frac{2\pi}{\beta(n_f)\log(\mu/\Lambda^{(n_f)})}
\end{equation}
with $\beta(n_f)\equiv 11 -2n_f/3$, and
\begin{align}
	\alpha_s^{(3)}(\mu) = \alpha_s^{(4)}(\mu,\delta m_c^\mathrm{sea})
\end{align}
at $\mu=m_c+\delta m_c^\mathrm{sea}$.
Here $\Lambda^{(3)}$ must be independent of $\delta m_c^\mathrm{sea}$, 
by decoupling, while $\Lambda^{(4)}$ must vary with $\delta m_c^\mathrm{sea}$
to cancel the effect of the shift in the match 
point $\mu=m_c+\delta m_c^\mathrm{sea}$.
It is straightforward to show that
\begin{align}
	\Lambda^{(4)}(\delta m_c^\mathrm{sea}) 
	&\approx m_c \left(\frac{\Lambda^{(3)}}{m_c}\right)^{\beta(3)/\beta(4)} 
	\left(1 - \frac{2}{25} \frac{\delta m_c^\mathrm{sea}}{m_c}\right)
	\nonumber \\
	&\approx \Lambda^{(4)}_\mathrm{phys} \times 
	\left(1 - \frac{2}{25} \frac{\delta m_c^\mathrm{sea}}{m_c}\right)
\end{align}
where $\Lambda^{(4)}_\mathrm{phys}$ is the value for physical sea-quark masses.
Thus the decoupling theorem requires that
\begin{equation}
	\alpha_s^{(4)}(\mu,\delta m_c^\mathrm{sea})
	= \alpha_s^{(4)}\left(\mu\times \left(1 + \frac{2}{25} 
	\frac{\delta m_c^\mathrm{sea}}{m_c}\right)\right).
\end{equation}
By comparing with Eqs.~(\ref{eq:alpha-tilde}) and~(\ref{eq:xi-alpha}), we see
that
\begin{equation}
	g_{c,\alpha} = \frac{2}{25} + \order(\alpha_s),
\end{equation}
and, therefore, that 
the lattice spacing varies with~$\delta m_c^\mathrm{sea}$ 
(Eq.~~(\ref{eq:a-msea})). 

There is an analogous effect in the heavy-quark mass, but the mass
dependence in~$\xi_m$ is suppressed by $\alpha_s^2$ and so is 
negligible in our analysis.

This analysis shows that a constant lattice spacing is incompatible with
the decoupling theorem. The scheme we use avoids this problem by allowing
the lattice spacing to vary with~$\delta m_c^\mathrm{sea}$, while 
holding the value of $w_0$ constant (as required by the decoupling theorem
applied to~$w_0$ itself). The violation of the decoupling theorem in the former
case is only apparent; results from all schemes should agree when 
the sea-quark masses are tuned to their physical values.

\section{Previous Method}
\label{app-previous}

\begin{table}
\caption{Simulations results for 
$\eta_h$ masses and reduced moments $R_n$ (old definition) with various bare
heavy-quark masses~$am_{0h}$ and gluon ensembles (first column,
see Table~\ref{tab:ensembles}). Data from gluon ensembles~1--3 are 
not listed because they were not used in the analysis in Appendix~\ref{app-previous}.
}
\label{tab:metah-oldRn}
	\begin{ruledtabular}
	\begin{tabular}{cclcccc}
& $am_{0h}$ & $\quad am_{\eta_h}$ & $R_4$ & $R_6$ & $R_8$ & $R_{10}$ \\ \hline
4&   0.645&    1.83976(11)&      1.1842(2)&      1.4857(2)&      1.3785(1)
&      1.3179(1)\\
&   0.663&    1.87456(12)&      1.1783(2)&      1.4755(2)&      1.3732(1)
&      1.3148(1)\\
 5&   0.627&     1.80318(8)&      1.1896(1)&      1.4944(1)&      1.3825(1)
 &      1.3201(1)\\
&   0.650&     1.84797(8)&      1.1819(1)&      1.4813(1)&      1.3759(1)
&      1.3162(1)\\
&   0.800&     2.13055(7)&      1.1409(1)&      1.4012(1)&      1.3304(1)
&      1.2880(1)\\
 6&   0.637&     1.82225(5)&      1.1860(1)&      1.4882(1)&      1.3793(1)
 &      1.3181(0)\\
\hline  7&   0.439&     1.34246(4)&      1.2134(1)&      1.5122(1)&      1.3758(1)
&      1.3089(0)\\
&   0.500&     1.47051(4)&      1.1886(1)&      1.4782(1)&      1.3586(1)
&      1.2968(0)\\
&   0.600&     1.67455(4)&      1.1565(1)&      1.4282(1)&      1.3334(0)
&      1.2801(0)\\
&   0.700&     1.87210(4)&      1.1315(0)&      1.3827(0)&      1.3089(0)
&      1.2647(0)\\
&   0.800&     2.06328(3)&      1.1118(0)&      1.3401(0)&      1.2834(0)
&      1.2482(0)\\
 8&   0.433&     1.32929(3)&      1.2160(1)&      1.5153(1)&      1.3772(0)
 &      1.3099(0)\\
&   0.500&     1.47012(3)&      1.1885(0)&      1.4777(1)&      1.3582(0)
&      1.2965(0)\\
&   0.600&     1.67418(3)&      1.1564(0)&      1.4279(0)&      1.3331(0)
&      1.2799(0)\\
&   0.700&     1.87177(2)&      1.1315(0)&      1.3824(0)&      1.3087(0)
&      1.2645(0)\\
&   0.800&     2.06297(2)&      1.1117(0)&      1.3399(0)&      1.2832(0)
&      1.2480(0)\\
\hline  9&   0.269&     0.88525(5)&      1.2401(4)&      1.5182(4)&      1.3711(2)
&      1.3046(2)\\
&   0.274&     0.89669(5)&      1.2368(4)&      1.5139(3)&      1.3686(2)
&      1.3028(1)\\
&   0.400&     1.17560(5)&      1.1752(2)&      1.4312(2)&      1.3199(1)
&      1.2660(1)\\
&   0.500&     1.38750(4)&      1.1440(2)&      1.3854(2)&      1.2943(1)
&      1.2465(1)\\
&   0.600&     1.59311(4)&      1.1204(1)&      1.3464(1)&      1.2734(1)
&      1.2316(1)\\
&   0.700&     1.79313(4)&      1.1018(1)&      1.3107(1)&      1.2535(1)
&      1.2183(1)\\
&   0.800&     1.98751(3)&      1.0867(1)&      1.2771(1)&      1.2328(0)
&      1.2046(0)\\
	\end{tabular}
	\end{ruledtabular}
\end{table}

The analysis in our previous ($n_f=3$) paper used a different definition
for the reduced moments with $n\ge6$:
\begin{equation}
	R_{n\ge6} = \frac{m_{\eta_h}}{2m_{0h}} \left(G_n/G_n^{(0)}\right)^{1/(n-4)}
\end{equation}
instead of \eq{rn-def}. As a result these moments 
equal $z(m_{\eta_h},\mu) \, r_n(\almsb,\mu)$ in perturbation theory 
where
\begin{equation}
	z(m_{\eta_c},\mu) \equiv \frac{m_{\eta_h}}{2m_h(\mu)}
\end{equation}
replaces $z_c(\mu)$, which is defined at the $c$~mass instead of $m_h$.
Fits to these moments give both the coupling and the function
$z(m_{\eta_h},\mu)$, from which the~$c$ and $b$~masses can be extracted.

We analyzed our data using the old definition, parameterizing the
$m_{\eta_h}$ dependence of $z(m_{\eta_c},\mu)$ with a cubic spline. 
The values for the $R_n$ moments used 
are given in Table~\ref{tab:metah-oldRn}. We obtained 
results that agree with the results obtained from our new method
to within a standard deviation, but are not quite as accurate:
\begin{align}
	\almsb(5\,\mathrm{GeV}, n_f=4) &= 0.2148(29) \\
	m_c(3\,\mathrm{GeV}, n_f=4) &= 0.9896(69).
\end{align}
The older method is more complicated because it attempts to determine
the coupling at the same time as it determines the functional 
dependence of $z(m_{\eta_h},\mu=3m_h)$. In the new method, 
$z(m_{\eta_h},\mu=3m_h)$ is replaced by~$z_c(\mu)$, whose  
dependence on $\mu$ is known \emph{a priori} from perturbative QCD.

\bibliography{paper}{}

\begin{thebibliography}{56}%
\makeatletter
\providecommand \@ifxundefined [1]{%
 \@ifx{#1\undefined}
}%
\providecommand \@ifnum [1]{%
 \ifnum #1\expandafter \@firstoftwo
 \else \expandafter \@secondoftwo
 \fi
}%
\providecommand \@ifx [1]{%
 \ifx #1\expandafter \@firstoftwo
 \else \expandafter \@secondoftwo
 \fi
}%
\providecommand \natexlab [1]{#1}%
\providecommand \enquote  [1]{``#1''}%
\providecommand \bibnamefont  [1]{#1}%
\providecommand \bibfnamefont [1]{#1}%
\providecommand \citenamefont [1]{#1}%
\providecommand \href@noop [0]{\@secondoftwo}%
\providecommand \href [0]{\begingroup \@sanitize@url \@href}%
\providecommand \@href[1]{\@@startlink{#1}\@@href}%
\providecommand \@@href[1]{\endgroup#1\@@endlink}%
\providecommand \@sanitize@url [0]{\catcode `\\12\catcode `\$12\catcode
  `\&12\catcode `\#12\catcode `\^12\catcode `\_12\catcode `\%12\relax}%
\providecommand \@@startlink[1]{}%
\providecommand \@@endlink[0]{}%
\providecommand \url  [0]{\begingroup\@sanitize@url \@url }%
\providecommand \@url [1]{\endgroup\@href {#1}{\urlprefix }}%
\providecommand \urlprefix  [0]{URL }%
\providecommand \Eprint [0]{\href }%
\providecommand \doibase [0]{http://dx.doi.org/}%
\providecommand \selectlanguage [0]{\@gobble}%
\providecommand \bibinfo  [0]{\@secondoftwo}%
\providecommand \bibfield  [0]{\@secondoftwo}%
\providecommand \translation [1]{[#1]}%
\providecommand \BibitemOpen [0]{}%
\providecommand \bibitemStop [0]{}%
\providecommand \bibitemNoStop [0]{.\EOS\space}%
\providecommand \EOS [0]{\spacefactor3000\relax}%
\providecommand \BibitemShut  [1]{\csname bibitem#1\endcsname}%
\let\auto@bib@innerbib\@empty
\bibitem [{\citenamefont {Lepage}\ \emph {et~al.}(2014)\citenamefont {Lepage},
  \citenamefont {Mackenzie},\ and\ \citenamefont {Peskin}}]{Lepage:2014fla}%
  \BibitemOpen
  \bibfield  {author} {\bibinfo {author} {\bibfnamefont {G.~P.}\ \bibnamefont
  {Lepage}}, \bibinfo {author} {\bibfnamefont {P.~B.}\ \bibnamefont
  {Mackenzie}}, \ and\ \bibinfo {author} {\bibfnamefont {M.~E.}\ \bibnamefont
  {Peskin}},\ }\href@noop {} {\  (\bibinfo {year} {2014})},\ \Eprint
  {http://arxiv.org/abs/1404.0319} {arXiv:1404.0319 [hep-ph]} \BibitemShut
  {NoStop}%
\bibitem [{\citenamefont {McNeile}\ \emph {et~al.}(2010)\citenamefont
  {McNeile}, \citenamefont {Davies}, \citenamefont {Follana}, \citenamefont
  {Hornbostel},\ and\ \citenamefont {Lepage}}]{McNeile:2010ji}%
  \BibitemOpen
  \bibfield  {author} {\bibinfo {author} {\bibfnamefont {C.}~\bibnamefont
  {McNeile}}, \bibinfo {author} {\bibfnamefont {C.}~\bibnamefont {Davies}},
  \bibinfo {author} {\bibfnamefont {E.}~\bibnamefont {Follana}}, \bibinfo
  {author} {\bibfnamefont {K.}~\bibnamefont {Hornbostel}}, \ and\ \bibinfo
  {author} {\bibfnamefont {G.}~\bibnamefont {Lepage}} (\bibinfo {collaboration}
  {HPQCD Collaboration}),\ }\href {\doibase 10.1103/PhysRevD.82.034512}
  {\bibfield  {journal} {\bibinfo  {journal} {Phys.Rev.}\ }\textbf {\bibinfo
  {volume} {D82}},\ \bibinfo {pages} {034512} (\bibinfo {year} {2010})},\
  \Eprint {http://arxiv.org/abs/1004.4285} {arXiv:1004.4285 [hep-lat]}
  \BibitemShut {NoStop}%
\bibitem [{\citenamefont {Follana}\ \emph {et~al.}(2007)\citenamefont {Follana}
  \emph {et~al.}}]{Follana:2006rc}%
  \BibitemOpen
  \bibfield  {author} {\bibinfo {author} {\bibfnamefont {E.}~\bibnamefont
  {Follana}} \emph {et~al.} (\bibinfo {collaboration} {HPQCD Collaboration,
  UKQCD Collaboration}),\ }\href {\doibase 10.1103/PhysRevD.75.054502}
  {\bibfield  {journal} {\bibinfo  {journal} {Phys.Rev.}\ }\textbf {\bibinfo
  {volume} {D75}},\ \bibinfo {pages} {054502} (\bibinfo {year} {2007})},\
  \Eprint {http://arxiv.org/abs/hep-lat/0610092} {arXiv:hep-lat/0610092
  [hep-lat]} \BibitemShut {NoStop}%
\bibitem [{\citenamefont {Hart}\ \emph {et~al.}(2009)\citenamefont {Hart},
  \citenamefont {von Hippel},\ and\ \citenamefont {Horgan}}]{Hart:2008sq}%
  \BibitemOpen
  \bibfield  {author} {\bibinfo {author} {\bibfnamefont {A.}~\bibnamefont
  {Hart}}, \bibinfo {author} {\bibfnamefont {G.}~\bibnamefont {von Hippel}}, \
  and\ \bibinfo {author} {\bibfnamefont {R.}~\bibnamefont {Horgan}} (\bibinfo
  {collaboration} {HPQCD Collaboration}),\ }\href {\doibase
  10.1103/PhysRevD.79.074008} {\bibfield  {journal} {\bibinfo  {journal}
  {Phys.Rev.}\ }\textbf {\bibinfo {volume} {D79}},\ \bibinfo {pages} {074008}
  (\bibinfo {year} {2009})},\ \Eprint {http://arxiv.org/abs/0812.0503}
  {arXiv:0812.0503 [hep-lat]} \BibitemShut {NoStop}%
\bibitem [{\citenamefont {Shifman}\ \emph {et~al.}(1979)\citenamefont
  {Shifman}, \citenamefont {Vainshtein},\ and\ \citenamefont
  {Zakharov}}]{Shifman:1978bx}%
  \BibitemOpen
  \bibfield  {author} {\bibinfo {author} {\bibfnamefont {M.~A.}\ \bibnamefont
  {Shifman}}, \bibinfo {author} {\bibfnamefont {A.}~\bibnamefont {Vainshtein}},
  \ and\ \bibinfo {author} {\bibfnamefont {V.~I.}\ \bibnamefont {Zakharov}},\
  }\href {\doibase 10.1016/0550-3213(79)90022-1} {\bibfield  {journal}
  {\bibinfo  {journal} {Nucl.Phys.}\ }\textbf {\bibinfo {volume} {B147}},\
  \bibinfo {pages} {385} (\bibinfo {year} {1979})}\BibitemShut {NoStop}%
\bibitem [{\citenamefont {Chetyrkin}\ \emph {et~al.}(2006)\citenamefont
  {Chetyrkin}, \citenamefont {Kuhn},\ and\ \citenamefont
  {Sturm}}]{Chetyrkin:2006xg}%
  \BibitemOpen
  \bibfield  {author} {\bibinfo {author} {\bibfnamefont {K.}~\bibnamefont
  {Chetyrkin}}, \bibinfo {author} {\bibfnamefont {J.~H.}\ \bibnamefont {Kuhn}},
  \ and\ \bibinfo {author} {\bibfnamefont {C.}~\bibnamefont {Sturm}},\ }\href
  {\doibase 10.1140/epjc/s2006-02610-y} {\bibfield  {journal} {\bibinfo
  {journal} {Eur.Phys.J.}\ }\textbf {\bibinfo {volume} {C48}},\ \bibinfo
  {pages} {107} (\bibinfo {year} {2006})},\ \Eprint
  {http://arxiv.org/abs/hep-ph/0604234} {arXiv:hep-ph/0604234 [hep-ph]}
  \BibitemShut {NoStop}%
\bibitem [{\citenamefont {Boughezal}\ \emph {et~al.}(2006)\citenamefont
  {Boughezal}, \citenamefont {Czakon},\ and\ \citenamefont
  {Schutzmeier}}]{Boughezal:2006px}%
  \BibitemOpen
  \bibfield  {author} {\bibinfo {author} {\bibfnamefont {R.}~\bibnamefont
  {Boughezal}}, \bibinfo {author} {\bibfnamefont {M.}~\bibnamefont {Czakon}}, \
  and\ \bibinfo {author} {\bibfnamefont {T.}~\bibnamefont {Schutzmeier}},\
  }\href {\doibase 10.1103/PhysRevD.74.074006} {\bibfield  {journal} {\bibinfo
  {journal} {Phys.Rev.}\ }\textbf {\bibinfo {volume} {D74}},\ \bibinfo {pages}
  {074006} (\bibinfo {year} {2006})},\ \Eprint
  {http://arxiv.org/abs/hep-ph/0605023} {arXiv:hep-ph/0605023 [hep-ph]}
  \BibitemShut {NoStop}%
\bibitem [{\citenamefont {Maier}\ \emph {et~al.}(2008)\citenamefont {Maier},
  \citenamefont {Maierhofer},\ and\ \citenamefont {Marqaurd}}]{Maier:2008he}%
  \BibitemOpen
  \bibfield  {author} {\bibinfo {author} {\bibfnamefont {A.}~\bibnamefont
  {Maier}}, \bibinfo {author} {\bibfnamefont {P.}~\bibnamefont {Maierhofer}}, \
  and\ \bibinfo {author} {\bibfnamefont {P.}~\bibnamefont {Marqaurd}},\ }\href
  {\doibase 10.1016/j.physletb.2008.09.041} {\bibfield  {journal} {\bibinfo
  {journal} {Phys.Lett.}\ }\textbf {\bibinfo {volume} {B669}},\ \bibinfo
  {pages} {88} (\bibinfo {year} {2008})},\ \Eprint
  {http://arxiv.org/abs/0806.3405} {arXiv:0806.3405 [hep-ph]} \BibitemShut
  {NoStop}%
\bibitem [{\citenamefont {Kiyo}\ \emph {et~al.}(2009)\citenamefont {Kiyo},
  \citenamefont {Maier}, \citenamefont {Maierhofer},\ and\ \citenamefont
  {Marquard}}]{Kiyo:2009gb}%
  \BibitemOpen
  \bibfield  {author} {\bibinfo {author} {\bibfnamefont {Y.}~\bibnamefont
  {Kiyo}}, \bibinfo {author} {\bibfnamefont {A.}~\bibnamefont {Maier}},
  \bibinfo {author} {\bibfnamefont {P.}~\bibnamefont {Maierhofer}}, \ and\
  \bibinfo {author} {\bibfnamefont {P.}~\bibnamefont {Marquard}},\ }\href
  {\doibase 10.1016/j.nuclphysb.2009.08.010} {\bibfield  {journal} {\bibinfo
  {journal} {Nucl.Phys.}\ }\textbf {\bibinfo {volume} {B823}},\ \bibinfo
  {pages} {269} (\bibinfo {year} {2009})},\ \Eprint
  {http://arxiv.org/abs/0907.2120} {arXiv:0907.2120 [hep-ph]} \BibitemShut
  {NoStop}%
\bibitem [{\citenamefont {Maier}\ \emph {et~al.}(2010)\citenamefont {Maier},
  \citenamefont {Maierhofer}, \citenamefont {Marquard},\ and\ \citenamefont
  {Smirnov}}]{Maier:2009fz}%
  \BibitemOpen
  \bibfield  {author} {\bibinfo {author} {\bibfnamefont {A.}~\bibnamefont
  {Maier}}, \bibinfo {author} {\bibfnamefont {P.}~\bibnamefont {Maierhofer}},
  \bibinfo {author} {\bibfnamefont {P.}~\bibnamefont {Marquard}}, \ and\
  \bibinfo {author} {\bibfnamefont {A.}~\bibnamefont {Smirnov}},\ }\href
  {\doibase 10.1016/j.nuclphysb.2009.08.011} {\bibfield  {journal} {\bibinfo
  {journal} {Nucl.Phys.}\ }\textbf {\bibinfo {volume} {B824}},\ \bibinfo
  {pages} {1} (\bibinfo {year} {2010})},\ \Eprint
  {http://arxiv.org/abs/0907.2117} {arXiv:0907.2117 [hep-ph]} \BibitemShut
  {NoStop}%
\bibitem [{\citenamefont {Broadhurst}\ \emph {et~al.}(1994)\citenamefont
  {Broadhurst}, \citenamefont {Baikov}, \citenamefont {Ilyin}, \citenamefont
  {Fleischer}, \citenamefont {Tarasov} \emph {et~al.}}]{Broadhurst:1994qj}%
  \BibitemOpen
  \bibfield  {author} {\bibinfo {author} {\bibfnamefont {D.~J.}\ \bibnamefont
  {Broadhurst}}, \bibinfo {author} {\bibfnamefont {P.}~\bibnamefont {Baikov}},
  \bibinfo {author} {\bibfnamefont {V.}~\bibnamefont {Ilyin}}, \bibinfo
  {author} {\bibfnamefont {J.}~\bibnamefont {Fleischer}}, \bibinfo {author}
  {\bibfnamefont {O.}~\bibnamefont {Tarasov}},  \emph {et~al.},\ }\href
  {\doibase 10.1016/0370-2693(94)90524-X} {\bibfield  {journal} {\bibinfo
  {journal} {Phys.Lett.}\ }\textbf {\bibinfo {volume} {B329}},\ \bibinfo
  {pages} {103} (\bibinfo {year} {1994})},\ \Eprint
  {http://arxiv.org/abs/hep-ph/9403274} {arXiv:hep-ph/9403274 [hep-ph]}
  \BibitemShut {NoStop}%
\bibitem [{\citenamefont {Martinelli}\ \emph {et~al.}(1995)\citenamefont
  {Martinelli}, \citenamefont {Pittori}, \citenamefont {Sachrajda},
  \citenamefont {Testa},\ and\ \citenamefont {Vladikas}}]{Martinelli:1994ty}%
  \BibitemOpen
  \bibfield  {author} {\bibinfo {author} {\bibfnamefont {G.}~\bibnamefont
  {Martinelli}}, \bibinfo {author} {\bibfnamefont {C.}~\bibnamefont {Pittori}},
  \bibinfo {author} {\bibfnamefont {C.~T.}\ \bibnamefont {Sachrajda}}, \bibinfo
  {author} {\bibfnamefont {M.}~\bibnamefont {Testa}}, \ and\ \bibinfo {author}
  {\bibfnamefont {A.}~\bibnamefont {Vladikas}},\ }\href {\doibase
  10.1016/0550-3213(95)00126-D} {\bibfield  {journal} {\bibinfo  {journal}
  {Nucl.Phys.}\ }\textbf {\bibinfo {volume} {B445}},\ \bibinfo {pages} {81}
  (\bibinfo {year} {1995})},\ \Eprint {http://arxiv.org/abs/hep-lat/9411010}
  {arXiv:hep-lat/9411010 [hep-lat]} \BibitemShut {NoStop}%
\bibitem [{\citenamefont {Bazavov}\ \emph {et~al.}(2010)\citenamefont {Bazavov}
  \emph {et~al.}}]{Bazavov:2010ru}%
  \BibitemOpen
  \bibfield  {author} {\bibinfo {author} {\bibfnamefont {A.}~\bibnamefont
  {Bazavov}} \emph {et~al.} (\bibinfo {collaboration} {MILC collaboration}),\
  }\href {\doibase 10.1103/PhysRevD.82.074501} {\bibfield  {journal} {\bibinfo
  {journal} {Phys.Rev.}\ }\textbf {\bibinfo {volume} {D82}},\ \bibinfo {pages}
  {074501} (\bibinfo {year} {2010})},\ \Eprint {http://arxiv.org/abs/1004.0342}
  {arXiv:1004.0342 [hep-lat]} \BibitemShut {NoStop}%
\bibitem [{\citenamefont {Bazavov}\ \emph {et~al.}(2013)\citenamefont {Bazavov}
  \emph {et~al.}}]{Bazavov:2012xda}%
  \BibitemOpen
  \bibfield  {author} {\bibinfo {author} {\bibfnamefont {A.}~\bibnamefont
  {Bazavov}} \emph {et~al.} (\bibinfo {collaboration} {MILC Collaboration}),\
  }\href {\doibase 10.1103/PhysRevD.87.054505} {\bibfield  {journal} {\bibinfo
  {journal} {Phys.Rev.}\ }\textbf {\bibinfo {volume} {D87}},\ \bibinfo {pages}
  {054505} (\bibinfo {year} {2013})},\ \Eprint {http://arxiv.org/abs/1212.4768}
  {arXiv:1212.4768 [hep-lat]} \BibitemShut {NoStop}%
\bibitem [{\citenamefont {Lepage}\ \emph {et~al.}(2002)\citenamefont {Lepage},
  \citenamefont {Clark}, \citenamefont {Davies}, \citenamefont {Hornbostel},
  \citenamefont {Mackenzie} \emph {et~al.}}]{Lepage:2001ym}%
  \BibitemOpen
  \bibfield  {author} {\bibinfo {author} {\bibfnamefont {G.}~\bibnamefont
  {Lepage}}, \bibinfo {author} {\bibfnamefont {B.}~\bibnamefont {Clark}},
  \bibinfo {author} {\bibfnamefont {C.}~\bibnamefont {Davies}}, \bibinfo
  {author} {\bibfnamefont {K.}~\bibnamefont {Hornbostel}}, \bibinfo {author}
  {\bibfnamefont {P.}~\bibnamefont {Mackenzie}},  \emph {et~al.},\ }\href
  {\doibase 10.1016/S0920-5632(01)01638-3} {\bibfield  {journal} {\bibinfo
  {journal} {Nucl.Phys.Proc.Suppl.}\ }\textbf {\bibinfo {volume} {106}},\
  \bibinfo {pages} {12} (\bibinfo {year} {2002})},\ \Eprint
  {http://arxiv.org/abs/hep-lat/0110175} {arXiv:hep-lat/0110175 [hep-lat]}
  \BibitemShut {NoStop}%
\bibitem [{\citenamefont {Borsanyi}\ \emph {et~al.}(2012)\citenamefont
  {Borsanyi}, \citenamefont {Durr}, \citenamefont {Fodor}, \citenamefont
  {Hoelbling}, \citenamefont {Katz} \emph {et~al.}}]{Borsanyi:2012zs}%
  \BibitemOpen
  \bibfield  {author} {\bibinfo {author} {\bibfnamefont {S.}~\bibnamefont
  {Borsanyi}}, \bibinfo {author} {\bibfnamefont {S.}~\bibnamefont {Durr}},
  \bibinfo {author} {\bibfnamefont {Z.}~\bibnamefont {Fodor}}, \bibinfo
  {author} {\bibfnamefont {C.}~\bibnamefont {Hoelbling}}, \bibinfo {author}
  {\bibfnamefont {S.~D.}\ \bibnamefont {Katz}},  \emph {et~al.},\ }\href
  {\doibase 10.1007/JHEP09(2012)010} {\bibfield  {journal} {\bibinfo  {journal}
  {JHEP}\ }\textbf {\bibinfo {volume} {1209}},\ \bibinfo {pages} {010}
  (\bibinfo {year} {2012})},\ \Eprint {http://arxiv.org/abs/1203.4469}
  {arXiv:1203.4469 [hep-lat]} \BibitemShut {NoStop}%
\bibitem [{\citenamefont {Dowdall}\ \emph {et~al.}(2013)\citenamefont
  {Dowdall}, \citenamefont {Davies}, \citenamefont {Lepage},\ and\
  \citenamefont {McNeile}}]{Dowdall:2013rya}%
  \BibitemOpen
  \bibfield  {author} {\bibinfo {author} {\bibfnamefont {R.}~\bibnamefont
  {Dowdall}}, \bibinfo {author} {\bibfnamefont {C.}~\bibnamefont {Davies}},
  \bibinfo {author} {\bibfnamefont {G.}~\bibnamefont {Lepage}}, \ and\ \bibinfo
  {author} {\bibfnamefont {C.}~\bibnamefont {McNeile}} (\bibinfo
  {collaboration} {HPQCD Collaboration}),\ }\href {\doibase
  10.1103/PhysRevD.88.074504} {\bibfield  {journal} {\bibinfo  {journal}
  {Phys.Rev.}\ }\textbf {\bibinfo {volume} {D88}},\ \bibinfo {pages} {074504}
  (\bibinfo {year} {2013})},\ \Eprint {http://arxiv.org/abs/1303.1670}
  {arXiv:1303.1670 [hep-lat]} \BibitemShut {NoStop}%
\bibitem [{\citenamefont {van Ritbergen}\ \emph {et~al.}(1997)\citenamefont
  {van Ritbergen}, \citenamefont {Vermaseren},\ and\ \citenamefont
  {Larin}}]{vanRitbergen:1997va}%
  \BibitemOpen
  \bibfield  {author} {\bibinfo {author} {\bibfnamefont {T.}~\bibnamefont {van
  Ritbergen}}, \bibinfo {author} {\bibfnamefont {J.}~\bibnamefont
  {Vermaseren}}, \ and\ \bibinfo {author} {\bibfnamefont {S.}~\bibnamefont
  {Larin}},\ }\href {\doibase 10.1016/S0370-2693(97)00370-5} {\bibfield
  {journal} {\bibinfo  {journal} {Phys.Lett.}\ }\textbf {\bibinfo {volume}
  {B400}},\ \bibinfo {pages} {379} (\bibinfo {year} {1997})},\ \Eprint
  {http://arxiv.org/abs/hep-ph/9701390} {arXiv:hep-ph/9701390 [hep-ph]}
  \BibitemShut {NoStop}%
\bibitem [{\citenamefont {Czakon}(2005)}]{Czakon:2004bu}%
  \BibitemOpen
  \bibfield  {author} {\bibinfo {author} {\bibfnamefont {M.}~\bibnamefont
  {Czakon}},\ }\href {\doibase 10.1016/j.nuclphysb.2005.01.012} {\bibfield
  {journal} {\bibinfo  {journal} {Nucl.Phys.}\ }\textbf {\bibinfo {volume}
  {B710}},\ \bibinfo {pages} {485} (\bibinfo {year} {2005})},\ \Eprint
  {http://arxiv.org/abs/hep-ph/0411261} {arXiv:hep-ph/0411261 [hep-ph]}
  \BibitemShut {NoStop}%
\bibitem [{\citenamefont {Vermaseren}\ \emph {et~al.}(1997)\citenamefont
  {Vermaseren}, \citenamefont {Larin},\ and\ \citenamefont {van
  Ritbergen}}]{Vermaseren:1997fq}%
  \BibitemOpen
  \bibfield  {author} {\bibinfo {author} {\bibfnamefont {J.}~\bibnamefont
  {Vermaseren}}, \bibinfo {author} {\bibfnamefont {S.}~\bibnamefont {Larin}}, \
  and\ \bibinfo {author} {\bibfnamefont {T.}~\bibnamefont {van Ritbergen}},\
  }\href {\doibase 10.1016/S0370-2693(97)00660-6} {\bibfield  {journal}
  {\bibinfo  {journal} {Phys.Lett.}\ }\textbf {\bibinfo {volume} {B405}},\
  \bibinfo {pages} {327} (\bibinfo {year} {1997})},\ \Eprint
  {http://arxiv.org/abs/hep-ph/9703284} {arXiv:hep-ph/9703284 [hep-ph]}
  \BibitemShut {NoStop}%
\bibitem [{\citenamefont {Chetyrkin}(1997)}]{Chetyrkin:1997dh}%
  \BibitemOpen
  \bibfield  {author} {\bibinfo {author} {\bibfnamefont {K.}~\bibnamefont
  {Chetyrkin}},\ }\href {\doibase 10.1016/S0370-2693(97)00535-2} {\bibfield
  {journal} {\bibinfo  {journal} {Phys.Lett.}\ }\textbf {\bibinfo {volume}
  {B404}},\ \bibinfo {pages} {161} (\bibinfo {year} {1997})},\ \Eprint
  {http://arxiv.org/abs/hep-ph/9703278} {arXiv:hep-ph/9703278 [hep-ph]}
  \BibitemShut {NoStop}%
\bibitem [{\citenamefont {Novikov}\ \emph {et~al.}(1985)\citenamefont
  {Novikov}, \citenamefont {Shifman}, \citenamefont {Vainshtein},\ and\
  \citenamefont {Zakharov}}]{Novikov:1984rf}%
  \BibitemOpen
  \bibfield  {author} {\bibinfo {author} {\bibfnamefont {V.}~\bibnamefont
  {Novikov}}, \bibinfo {author} {\bibfnamefont {M.~A.}\ \bibnamefont
  {Shifman}}, \bibinfo {author} {\bibfnamefont {A.}~\bibnamefont {Vainshtein}},
  \ and\ \bibinfo {author} {\bibfnamefont {V.~I.}\ \bibnamefont {Zakharov}},\
  }\href {\doibase 10.1016/0550-3213(85)90087-2} {\bibfield  {journal}
  {\bibinfo  {journal} {Nucl.Phys.}\ }\textbf {\bibinfo {volume} {B249}},\
  \bibinfo {pages} {445} (\bibinfo {year} {1985})}\BibitemShut {NoStop}%
\bibitem [{\citenamefont {Shifman}(1998)}]{Shifman:1998rb}%
  \BibitemOpen
  \bibfield  {author} {\bibinfo {author} {\bibfnamefont {M.~A.}\ \bibnamefont
  {Shifman}},\ }\href {\doibase 10.1143/PTPS.131.1} {\bibfield  {journal}
  {\bibinfo  {journal} {Prog.Theor.Phys.Suppl.}\ }\textbf {\bibinfo {volume}
  {131}},\ \bibinfo {pages} {1} (\bibinfo {year} {1998})},\ \Eprint
  {http://arxiv.org/abs/hep-ph/9802214} {arXiv:hep-ph/9802214 [hep-ph]}
  \BibitemShut {NoStop}%
\bibitem [{\citenamefont {Shifman}(2013)}]{Shifman:2013uka}%
  \BibitemOpen
  \bibfield  {author} {\bibinfo {author} {\bibfnamefont {M.}~\bibnamefont
  {Shifman}},\ }\href@noop {} {\  (\bibinfo {year} {2013})},\ \Eprint
  {http://arxiv.org/abs/1310.1966} {arXiv:1310.1966 [hep-th]} \BibitemShut
  {NoStop}%
\bibitem [{\citenamefont {Hollands}\ and\ \citenamefont
  {Kopper}(2012)}]{Hollands:2011gf}%
  \BibitemOpen
  \bibfield  {author} {\bibinfo {author} {\bibfnamefont {S.}~\bibnamefont
  {Hollands}}\ and\ \bibinfo {author} {\bibfnamefont {C.}~\bibnamefont
  {Kopper}},\ }\href {\doibase 10.1007/s00220-012-1457-4} {\bibfield  {journal}
  {\bibinfo  {journal} {Commun.Math.Phys.}\ }\textbf {\bibinfo {volume}
  {313}},\ \bibinfo {pages} {257} (\bibinfo {year} {2012})},\ \Eprint
  {http://arxiv.org/abs/1105.3375} {arXiv:1105.3375 [hep-th]} \BibitemShut
  {NoStop}%
\bibitem [{\citenamefont {Pappadopulo}\ \emph {et~al.}(2012)\citenamefont
  {Pappadopulo}, \citenamefont {Rychkov}, \citenamefont {Espin},\ and\
  \citenamefont {Rattazzi}}]{Pappadopulo:2012jk}%
  \BibitemOpen
  \bibfield  {author} {\bibinfo {author} {\bibfnamefont {D.}~\bibnamefont
  {Pappadopulo}}, \bibinfo {author} {\bibfnamefont {S.}~\bibnamefont
  {Rychkov}}, \bibinfo {author} {\bibfnamefont {J.}~\bibnamefont {Espin}}, \
  and\ \bibinfo {author} {\bibfnamefont {R.}~\bibnamefont {Rattazzi}},\ }\href
  {\doibase 10.1103/PhysRevD.86.105043} {\bibfield  {journal} {\bibinfo
  {journal} {Phys.Rev.}\ }\textbf {\bibinfo {volume} {D86}},\ \bibinfo {pages}
  {105043} (\bibinfo {year} {2012})},\ \Eprint {http://arxiv.org/abs/1208.6449}
  {arXiv:1208.6449 [hep-th]} \BibitemShut {NoStop}%
\bibitem [{\citenamefont {Chetyrkin}\ \emph {et~al.}(1998)\citenamefont
  {Chetyrkin}, \citenamefont {Kniehl},\ and\ \citenamefont
  {Steinhauser}}]{Chetyrkin:1997un}%
  \BibitemOpen
  \bibfield  {author} {\bibinfo {author} {\bibfnamefont {K.}~\bibnamefont
  {Chetyrkin}}, \bibinfo {author} {\bibfnamefont {B.~A.}\ \bibnamefont
  {Kniehl}}, \ and\ \bibinfo {author} {\bibfnamefont {M.}~\bibnamefont
  {Steinhauser}},\ }\href {\doibase 10.1016/S0550-3213(97)00649-4} {\bibfield
  {journal} {\bibinfo  {journal} {Nucl.Phys.}\ }\textbf {\bibinfo {volume}
  {B510}},\ \bibinfo {pages} {61} (\bibinfo {year} {1998})},\ \Eprint
  {http://arxiv.org/abs/hep-ph/9708255} {arXiv:hep-ph/9708255 [hep-ph]}
  \BibitemShut {NoStop}%
\bibitem [{\citenamefont {Olive}\ \emph {et~al.}(2014)\citenamefont {Olive}
  \emph {et~al.}}]{Agashe:2014kda}%
  \BibitemOpen
  \bibfield  {author} {\bibinfo {author} {\bibfnamefont {K.}~\bibnamefont
  {Olive}} \emph {et~al.} (\bibinfo {collaboration} {Particle Data Group}),\
  }\href {\doibase 10.1088/1674-1137/38/9/090001} {\bibfield  {journal}
  {\bibinfo  {journal} {Chin.Phys.}\ }\textbf {\bibinfo {volume} {C38}},\
  \bibinfo {pages} {090001} (\bibinfo {year} {2014})}\BibitemShut {NoStop}%
\bibitem [{\citenamefont {McNeile}\ \emph {et~al.}(2013)\citenamefont
  {McNeile}, \citenamefont {Bazavov}, \citenamefont {Davies}, \citenamefont
  {Dowdall}, \citenamefont {Hornbostel} \emph {et~al.}}]{McNeile:2012xh}%
  \BibitemOpen
  \bibfield  {author} {\bibinfo {author} {\bibfnamefont {C.}~\bibnamefont
  {McNeile}}, \bibinfo {author} {\bibfnamefont {A.}~\bibnamefont {Bazavov}},
  \bibinfo {author} {\bibfnamefont {C.}~\bibnamefont {Davies}}, \bibinfo
  {author} {\bibfnamefont {R.}~\bibnamefont {Dowdall}}, \bibinfo {author}
  {\bibfnamefont {K.}~\bibnamefont {Hornbostel}},  \emph {et~al.},\ }\href
  {\doibase 10.1103/PhysRevD.87.034503} {\bibfield  {journal} {\bibinfo
  {journal} {Phys.Rev.}\ }\textbf {\bibinfo {volume} {D87}},\ \bibinfo {pages}
  {034503} (\bibinfo {year} {2013})},\ \Eprint {http://arxiv.org/abs/1211.6577}
  {arXiv:1211.6577 [hep-lat]} \BibitemShut {NoStop}%
\bibitem [{\citenamefont {Allison}\ \emph {et~al.}(2008)\citenamefont {Allison}
  \emph {et~al.}}]{Allison:2008xk}%
  \BibitemOpen
  \bibfield  {author} {\bibinfo {author} {\bibfnamefont {I.}~\bibnamefont
  {Allison}} \emph {et~al.},\ }\href {\doibase 10.1103/PhysRevD.78.054513}
  {\bibfield  {journal} {\bibinfo  {journal} {Phys.Rev.}\ }\textbf {\bibinfo
  {volume} {D78}},\ \bibinfo {pages} {054513} (\bibinfo {year} {2008})},\
  \Eprint {http://arxiv.org/abs/0805.2999} {arXiv:0805.2999 [hep-lat]}
  \BibitemShut {NoStop}%
\bibitem [{\citenamefont {Bouchard}\ \emph {et~al.}(2014)\citenamefont
  {Bouchard}, \citenamefont {Lepage}, \citenamefont {Monahan}, \citenamefont
  {Na},\ and\ \citenamefont {Shigemitsu}}]{Bouchard:2014ypa}%
  \BibitemOpen
  \bibfield  {author} {\bibinfo {author} {\bibfnamefont {C.}~\bibnamefont
  {Bouchard}}, \bibinfo {author} {\bibfnamefont {G.~P.}\ \bibnamefont
  {Lepage}}, \bibinfo {author} {\bibfnamefont {C.}~\bibnamefont {Monahan}},
  \bibinfo {author} {\bibfnamefont {H.}~\bibnamefont {Na}}, \ and\ \bibinfo
  {author} {\bibfnamefont {J.}~\bibnamefont {Shigemitsu}},\ }\href@noop {} {\
  (\bibinfo {year} {2014})},\ \Eprint {http://arxiv.org/abs/1406.2279}
  {arXiv:1406.2279 [hep-lat]} \BibitemShut {NoStop}%
\bibitem [{\citenamefont {Davies}\ \emph
  {et~al.}(2010{\natexlab{a}})\citenamefont {Davies}, \citenamefont {McNeile},
  \citenamefont {Wong}, \citenamefont {Follana}, \citenamefont {Horgan} \emph
  {et~al.}}]{Davies:2009ih}%
  \BibitemOpen
  \bibfield  {author} {\bibinfo {author} {\bibfnamefont {C.}~\bibnamefont
  {Davies}}, \bibinfo {author} {\bibfnamefont {C.}~\bibnamefont {McNeile}},
  \bibinfo {author} {\bibfnamefont {K.}~\bibnamefont {Wong}}, \bibinfo {author}
  {\bibfnamefont {E.}~\bibnamefont {Follana}}, \bibinfo {author} {\bibfnamefont
  {R.}~\bibnamefont {Horgan}},  \emph {et~al.} (\bibinfo {collaboration} {HPQCD
  Collaboration}),\ }\href {\doibase 10.1103/PhysRevLett.104.132003} {\bibfield
   {journal} {\bibinfo  {journal} {Phys.Rev.Lett.}\ }\textbf {\bibinfo {volume}
  {104}},\ \bibinfo {pages} {132003} (\bibinfo {year} {2010}{\natexlab{a}})},\
  \Eprint {http://arxiv.org/abs/0910.3102} {arXiv:0910.3102 [hep-ph]}
  \BibitemShut {NoStop}%
\bibitem [{\citenamefont {Bazavov}\ \emph {et~al.}(2014)\citenamefont {Bazavov}
  \emph {et~al.}}]{Bazavov:2014wgs}%
  \BibitemOpen
  \bibfield  {author} {\bibinfo {author} {\bibfnamefont {A.}~\bibnamefont
  {Bazavov}} \emph {et~al.} (\bibinfo {collaboration} {Fermilab Lattice and
  MILC Collaborations}),\ }\href@noop {} {\  (\bibinfo {year} {2014})},\
  \Eprint {http://arxiv.org/abs/1407.3772} {arXiv:1407.3772 [hep-lat]}
  \BibitemShut {NoStop}%
\bibitem [{\citenamefont {Chetyrkin}\ \emph {et~al.}(2009)\citenamefont
  {Chetyrkin}, \citenamefont {Kuhn}, \citenamefont {Maier}, \citenamefont
  {Maierhofer}, \citenamefont {Marquard} \emph {et~al.}}]{Chetyrkin:2009fv}%
  \BibitemOpen
  \bibfield  {author} {\bibinfo {author} {\bibfnamefont {K.}~\bibnamefont
  {Chetyrkin}}, \bibinfo {author} {\bibfnamefont {J.}~\bibnamefont {Kuhn}},
  \bibinfo {author} {\bibfnamefont {A.}~\bibnamefont {Maier}}, \bibinfo
  {author} {\bibfnamefont {P.}~\bibnamefont {Maierhofer}}, \bibinfo {author}
  {\bibfnamefont {P.}~\bibnamefont {Marquard}},  \emph {et~al.},\ }\href
  {\doibase 10.1103/PhysRevD.80.074010} {\bibfield  {journal} {\bibinfo
  {journal} {Phys.Rev.}\ }\textbf {\bibinfo {volume} {D80}},\ \bibinfo {pages}
  {074010} (\bibinfo {year} {2009})},\ \Eprint {http://arxiv.org/abs/0907.2110}
  {arXiv:0907.2110 [hep-ph]} \BibitemShut {NoStop}%
\bibitem [{\citenamefont {Chetyrkin}\ \emph {et~al.}(2012)\citenamefont
  {Chetyrkin}, \citenamefont {Kuhn}, \citenamefont {Maier}, \citenamefont
  {Maierhofer}, \citenamefont {Marquard} \emph {et~al.}}]{Chetyrkin:2010ic}%
  \BibitemOpen
  \bibfield  {author} {\bibinfo {author} {\bibfnamefont {K.}~\bibnamefont
  {Chetyrkin}}, \bibinfo {author} {\bibfnamefont {J.}~\bibnamefont {Kuhn}},
  \bibinfo {author} {\bibfnamefont {A.}~\bibnamefont {Maier}}, \bibinfo
  {author} {\bibfnamefont {P.}~\bibnamefont {Maierhofer}}, \bibinfo {author}
  {\bibfnamefont {P.}~\bibnamefont {Marquard}},  \emph {et~al.},\ }\href
  {\doibase 10.1007/s11232-012-0024-7} {\bibfield  {journal} {\bibinfo
  {journal} {Theor.Math.Phys.}\ }\textbf {\bibinfo {volume} {170}},\ \bibinfo
  {pages} {217} (\bibinfo {year} {2012})},\ \Eprint
  {http://arxiv.org/abs/1010.6157} {arXiv:1010.6157 [hep-ph]} \BibitemShut
  {NoStop}%
\bibitem [{\citenamefont {Carrasco}\ \emph {et~al.}(2014)\citenamefont
  {Carrasco}, \citenamefont {Deuzeman}, \citenamefont {Dimopoulos},
  \citenamefont {Frezzotti}, \citenamefont {Gimenez} \emph
  {et~al.}}]{Carrasco:2014cwa}%
  \BibitemOpen
  \bibfield  {author} {\bibinfo {author} {\bibfnamefont {N.}~\bibnamefont
  {Carrasco}}, \bibinfo {author} {\bibfnamefont {A.}~\bibnamefont {Deuzeman}},
  \bibinfo {author} {\bibfnamefont {P.}~\bibnamefont {Dimopoulos}}, \bibinfo
  {author} {\bibfnamefont {R.}~\bibnamefont {Frezzotti}}, \bibinfo {author}
  {\bibfnamefont {V.}~\bibnamefont {Gimenez}},  \emph {et~al.},\ }\href@noop {}
  {\  (\bibinfo {year} {2014})},\ \Eprint {http://arxiv.org/abs/1403.4504}
  {arXiv:1403.4504 [hep-lat]} \BibitemShut {NoStop}%
\bibitem [{\citenamefont {Durr}\ and\ \citenamefont
  {Koutsou}(2012)}]{Durr:2011ed}%
  \BibitemOpen
  \bibfield  {author} {\bibinfo {author} {\bibfnamefont {S.}~\bibnamefont
  {Durr}}\ and\ \bibinfo {author} {\bibfnamefont {G.}~\bibnamefont {Koutsou}},\
  }\href {\doibase 10.1103/PhysRevLett.108.122003} {\bibfield  {journal}
  {\bibinfo  {journal} {Phys.Rev.Lett.}\ }\textbf {\bibinfo {volume} {108}},\
  \bibinfo {pages} {122003} (\bibinfo {year} {2012})},\ \Eprint
  {http://arxiv.org/abs/1108.1650} {arXiv:1108.1650 [hep-lat]} \BibitemShut
  {NoStop}%
\bibitem [{\citenamefont {Blossier}\ \emph {et~al.}(2010)\citenamefont
  {Blossier} \emph {et~al.}}]{Blossier:2010cr}%
  \BibitemOpen
  \bibfield  {author} {\bibinfo {author} {\bibfnamefont {B.}~\bibnamefont
  {Blossier}} \emph {et~al.} (\bibinfo {collaboration} {ETM Collaboration}),\
  }\href {\doibase 10.1103/PhysRevD.82.114513} {\bibfield  {journal} {\bibinfo
  {journal} {Phys.Rev.}\ }\textbf {\bibinfo {volume} {D82}},\ \bibinfo {pages}
  {114513} (\bibinfo {year} {2010})},\ \Eprint {http://arxiv.org/abs/1010.3659}
  {arXiv:1010.3659 [hep-lat]} \BibitemShut {NoStop}%
\bibitem [{\citenamefont {Durr}\ \emph {et~al.}(2011)\citenamefont {Durr},
  \citenamefont {Fodor}, \citenamefont {Hoelbling}, \citenamefont {Katz},
  \citenamefont {Krieg} \emph {et~al.}}]{Durr:2010vn}%
  \BibitemOpen
  \bibfield  {author} {\bibinfo {author} {\bibfnamefont {S.}~\bibnamefont
  {Durr}}, \bibinfo {author} {\bibfnamefont {Z.}~\bibnamefont {Fodor}},
  \bibinfo {author} {\bibfnamefont {C.}~\bibnamefont {Hoelbling}}, \bibinfo
  {author} {\bibfnamefont {S.}~\bibnamefont {Katz}}, \bibinfo {author}
  {\bibfnamefont {S.}~\bibnamefont {Krieg}},  \emph {et~al.},\ }\href {\doibase
  10.1016/j.physletb.2011.05.053} {\bibfield  {journal} {\bibinfo  {journal}
  {Phys.Lett.}\ }\textbf {\bibinfo {volume} {B701}},\ \bibinfo {pages} {265}
  (\bibinfo {year} {2011})},\ \Eprint {http://arxiv.org/abs/1011.2403}
  {arXiv:1011.2403 [hep-lat]} \BibitemShut {NoStop}%
\bibitem [{\citenamefont {Arthur}\ \emph {et~al.}(2013)\citenamefont {Arthur}
  \emph {et~al.}}]{Arthur:2012opa}%
  \BibitemOpen
  \bibfield  {author} {\bibinfo {author} {\bibfnamefont {R.}~\bibnamefont
  {Arthur}} \emph {et~al.} (\bibinfo {collaboration} {RBC Collaboration, UKQCD
  Collaboration}),\ }\href {\doibase 10.1103/PhysRevD.87.094514} {\bibfield
  {journal} {\bibinfo  {journal} {Phys.Rev.}\ }\textbf {\bibinfo {volume}
  {D87}},\ \bibinfo {pages} {094514} (\bibinfo {year} {2013})},\ \Eprint
  {http://arxiv.org/abs/1208.4412} {arXiv:1208.4412 [hep-lat]} \BibitemShut
  {NoStop}%
\bibitem [{\citenamefont {Georgi}\ and\ \citenamefont
  {Jarlskog}(1979)}]{Georgi:1979df}%
  \BibitemOpen
  \bibfield  {author} {\bibinfo {author} {\bibfnamefont {H.}~\bibnamefont
  {Georgi}}\ and\ \bibinfo {author} {\bibfnamefont {C.}~\bibnamefont
  {Jarlskog}},\ }\href {\doibase 10.1016/0370-2693(79)90842-6} {\bibfield
  {journal} {\bibinfo  {journal} {Phys.Lett.}\ }\textbf {\bibinfo {volume}
  {B86}},\ \bibinfo {pages} {297} (\bibinfo {year} {1979})}\BibitemShut
  {NoStop}%
\bibitem [{\citenamefont {Davies}\ \emph {et~al.}(2008)\citenamefont {Davies}
  \emph {et~al.}}]{Davies:2008sw}%
  \BibitemOpen
  \bibfield  {author} {\bibinfo {author} {\bibfnamefont {C.}~\bibnamefont
  {Davies}} \emph {et~al.} (\bibinfo {collaboration} {HPQCD Collaboration}),\
  }\href {\doibase 10.1103/PhysRevD.78.114507} {\bibfield  {journal} {\bibinfo
  {journal} {Phys.Rev.}\ }\textbf {\bibinfo {volume} {D78}},\ \bibinfo {pages}
  {114507} (\bibinfo {year} {2008})},\ \Eprint {http://arxiv.org/abs/0807.1687}
  {arXiv:0807.1687 [hep-lat]} \BibitemShut {NoStop}%
\bibitem [{\citenamefont {Shintani}\ \emph {et~al.}(2010)\citenamefont
  {Shintani}, \citenamefont {Aoki}, \citenamefont {Fukaya}, \citenamefont
  {Hashimoto}, \citenamefont {Kaneko} \emph {et~al.}}]{Shintani:2010ph}%
  \BibitemOpen
  \bibfield  {author} {\bibinfo {author} {\bibfnamefont {E.}~\bibnamefont
  {Shintani}}, \bibinfo {author} {\bibfnamefont {S.}~\bibnamefont {Aoki}},
  \bibinfo {author} {\bibfnamefont {H.}~\bibnamefont {Fukaya}}, \bibinfo
  {author} {\bibfnamefont {S.}~\bibnamefont {Hashimoto}}, \bibinfo {author}
  {\bibfnamefont {T.}~\bibnamefont {Kaneko}},  \emph {et~al.},\ }\href
  {\doibase 10.1103/PhysRevD.82.074505;10.1103/PhysRevD.89.099903,
  10.1103/PhysRevD.82.074505} {\bibfield  {journal} {\bibinfo  {journal}
  {Phys.Rev.}\ }\textbf {\bibinfo {volume} {D82}},\ \bibinfo {pages} {074505}
  (\bibinfo {year} {2010})},\ \Eprint {http://arxiv.org/abs/1002.0371}
  {arXiv:1002.0371 [hep-lat]} \BibitemShut {NoStop}%
\bibitem [{\citenamefont {Bazavov}\ \emph {et~al.}(2012)\citenamefont
  {Bazavov}, \citenamefont {Brambilla}, \citenamefont {Garcia~i Tormo},
  \citenamefont {Petreczky}, \citenamefont {Soto} \emph
  {et~al.}}]{Bazavov:2012ka}%
  \BibitemOpen
  \bibfield  {author} {\bibinfo {author} {\bibfnamefont {A.}~\bibnamefont
  {Bazavov}}, \bibinfo {author} {\bibfnamefont {N.}~\bibnamefont {Brambilla}},
  \bibinfo {author} {\bibfnamefont {X.}~\bibnamefont {Garcia~i Tormo}},
  \bibinfo {author} {\bibfnamefont {P.}~\bibnamefont {Petreczky}}, \bibinfo
  {author} {\bibfnamefont {J.}~\bibnamefont {Soto}},  \emph {et~al.},\ }\href
  {\doibase 10.1103/PhysRevD.86.114031} {\bibfield  {journal} {\bibinfo
  {journal} {Phys.Rev.}\ }\textbf {\bibinfo {volume} {D86}},\ \bibinfo {pages}
  {114031} (\bibinfo {year} {2012})},\ \Eprint {http://arxiv.org/abs/1205.6155}
  {arXiv:1205.6155 [hep-ph]} \BibitemShut {NoStop}%
\bibitem [{\citenamefont {Jansen}\ \emph {et~al.}(2012)\citenamefont {Jansen},
  \citenamefont {Karbstein}, \citenamefont {Nagy},\ and\ \citenamefont
  {Wagner}}]{Jansen:2011vv}%
  \BibitemOpen
  \bibfield  {author} {\bibinfo {author} {\bibfnamefont {K.}~\bibnamefont
  {Jansen}}, \bibinfo {author} {\bibfnamefont {F.}~\bibnamefont {Karbstein}},
  \bibinfo {author} {\bibfnamefont {A.}~\bibnamefont {Nagy}}, \ and\ \bibinfo
  {author} {\bibfnamefont {M.}~\bibnamefont {Wagner}} (\bibinfo {collaboration}
  {ETM Collaboration}),\ }\href {\doibase 10.1007/JHEP01(2012)025} {\bibfield
  {journal} {\bibinfo  {journal} {JHEP}\ }\textbf {\bibinfo {volume} {1201}},\
  \bibinfo {pages} {025} (\bibinfo {year} {2012})},\ \Eprint
  {http://arxiv.org/abs/1110.6859} {arXiv:1110.6859 [hep-ph]} \BibitemShut
  {NoStop}%
\bibitem [{\citenamefont {Fritzsch}\ \emph {et~al.}(2012)\citenamefont
  {Fritzsch}, \citenamefont {Knechtli}, \citenamefont {Leder}, \citenamefont
  {Marinkovic}, \citenamefont {Schaefer} \emph {et~al.}}]{Fritzsch:2012wq}%
  \BibitemOpen
  \bibfield  {author} {\bibinfo {author} {\bibfnamefont {P.}~\bibnamefont
  {Fritzsch}}, \bibinfo {author} {\bibfnamefont {F.}~\bibnamefont {Knechtli}},
  \bibinfo {author} {\bibfnamefont {B.}~\bibnamefont {Leder}}, \bibinfo
  {author} {\bibfnamefont {M.}~\bibnamefont {Marinkovic}}, \bibinfo {author}
  {\bibfnamefont {S.}~\bibnamefont {Schaefer}},  \emph {et~al.},\ }\href
  {\doibase 10.1016/j.nuclphysb.2012.07.026} {\bibfield  {journal} {\bibinfo
  {journal} {Nucl.Phys.}\ }\textbf {\bibinfo {volume} {B865}},\ \bibinfo
  {pages} {397} (\bibinfo {year} {2012})},\ \Eprint
  {http://arxiv.org/abs/1205.5380} {arXiv:1205.5380 [hep-lat]} \BibitemShut
  {NoStop}%
\bibitem [{\citenamefont {Blossier}\ \emph {et~al.}(2014)\citenamefont
  {Blossier} \emph {et~al.}}]{Blossier:2013ioa}%
  \BibitemOpen
  \bibfield  {author} {\bibinfo {author} {\bibfnamefont {B.}~\bibnamefont
  {Blossier}} \emph {et~al.} (\bibinfo {collaboration} {ETM Collaboration}),\
  }\href {\doibase 10.1103/PhysRevD.89.014507} {\bibfield  {journal} {\bibinfo
  {journal} {Phys.Rev.}\ }\textbf {\bibinfo {volume} {D89}},\ \bibinfo {pages}
  {014507} (\bibinfo {year} {2014})},\ \Eprint {http://arxiv.org/abs/1310.3763}
  {arXiv:1310.3763 [hep-ph]} \BibitemShut {NoStop}%
\bibitem [{\citenamefont {Lee}\ \emph {et~al.}(2013)\citenamefont {Lee} \emph
  {et~al.}}]{Lee:2013mla}%
  \BibitemOpen
  \bibfield  {author} {\bibinfo {author} {\bibfnamefont {A.}~\bibnamefont
  {Lee}} \emph {et~al.} (\bibinfo {collaboration} {HPQCD Collaboration}),\
  }\href {\doibase 10.1103/PhysRevD.87.074018} {\bibfield  {journal} {\bibinfo
  {journal} {Phys.Rev.}\ }\textbf {\bibinfo {volume} {D87}},\ \bibinfo {pages}
  {074018} (\bibinfo {year} {2013})},\ \Eprint {http://arxiv.org/abs/1302.3739}
  {arXiv:1302.3739 [hep-lat]} \BibitemShut {NoStop}%
\bibitem [{\citenamefont {El-Khadra}\ \emph {et~al.}(1997)\citenamefont
  {El-Khadra}, \citenamefont {Kronfeld},\ and\ \citenamefont
  {Mackenzie}}]{ElKhadra:1996mp}%
  \BibitemOpen
  \bibfield  {author} {\bibinfo {author} {\bibfnamefont {A.~X.}\ \bibnamefont
  {El-Khadra}}, \bibinfo {author} {\bibfnamefont {A.~S.}\ \bibnamefont
  {Kronfeld}}, \ and\ \bibinfo {author} {\bibfnamefont {P.~B.}\ \bibnamefont
  {Mackenzie}},\ }\href {\doibase 10.1103/PhysRevD.55.3933} {\bibfield
  {journal} {\bibinfo  {journal} {Phys.Rev.}\ }\textbf {\bibinfo {volume}
  {D55}},\ \bibinfo {pages} {3933} (\bibinfo {year} {1997})},\ \Eprint
  {http://arxiv.org/abs/hep-lat/9604004} {arXiv:hep-lat/9604004 [hep-lat]}
  \BibitemShut {NoStop}%
\bibitem [{\citenamefont {Colquhoun}\ \emph {et~al.}(2014)\citenamefont
  {Colquhoun}, \citenamefont {Dowdall}, \citenamefont {Davies}, \citenamefont
  {Hornbostel},\ and\ \citenamefont {Lepage}}]{Colquhoun:2014ica}%
  \BibitemOpen
  \bibfield  {author} {\bibinfo {author} {\bibfnamefont {B.}~\bibnamefont
  {Colquhoun}}, \bibinfo {author} {\bibfnamefont {R.}~\bibnamefont {Dowdall}},
  \bibinfo {author} {\bibfnamefont {C.}~\bibnamefont {Davies}}, \bibinfo
  {author} {\bibfnamefont {K.}~\bibnamefont {Hornbostel}}, \ and\ \bibinfo
  {author} {\bibfnamefont {G.}~\bibnamefont {Lepage}} (\bibinfo {collaboration}
  {HPQCD Collaboration}),\ }\href@noop {} {\  (\bibinfo {year} {2014})},\
  \Eprint {http://arxiv.org/abs/1408.5768} {arXiv:1408.5768 [hep-lat]}
  \BibitemShut {NoStop}%
\bibitem [{\citenamefont {Luscher}(2010)}]{Luscher:2010iy}%
  \BibitemOpen
  \bibfield  {author} {\bibinfo {author} {\bibfnamefont {M.}~\bibnamefont
  {Luscher}},\ }\href {\doibase 10.1007/JHEP08(2010)071} {\bibfield  {journal}
  {\bibinfo  {journal} {JHEP}\ }\textbf {\bibinfo {volume} {1008}},\ \bibinfo
  {pages} {071} (\bibinfo {year} {2010})},\ \Eprint
  {http://arxiv.org/abs/1006.4518} {arXiv:1006.4518 [hep-lat]} \BibitemShut
  {NoStop}%
\bibitem [{\citenamefont {B{\"a}r}\ and\ \citenamefont
  {Golterman}(2014)}]{Bar:2013ora}%
  \BibitemOpen
  \bibfield  {author} {\bibinfo {author} {\bibfnamefont {O.}~\bibnamefont
  {B{\"a}r}}\ and\ \bibinfo {author} {\bibfnamefont {M.}~\bibnamefont
  {Golterman}},\ }\href {\doibase 10.1103/PhysRevD.89.034505} {\bibfield
  {journal} {\bibinfo  {journal} {Phys.Rev.}\ }\textbf {\bibinfo {volume}
  {D89}},\ \bibinfo {pages} {034505} (\bibinfo {year} {2014})},\ \Eprint
  {http://arxiv.org/abs/1312.4999} {arXiv:1312.4999 [hep-lat]} \BibitemShut
  {NoStop}%
\bibitem [{\citenamefont {Davies}\ \emph
  {et~al.}(2010{\natexlab{b}})\citenamefont {Davies}, \citenamefont {McNeile},
  \citenamefont {Follana}, \citenamefont {Lepage}, \citenamefont {Na} \emph
  {et~al.}}]{fdsupdate}%
  \BibitemOpen
  \bibfield  {author} {\bibinfo {author} {\bibfnamefont {C.}~\bibnamefont
  {Davies}}, \bibinfo {author} {\bibfnamefont {C.}~\bibnamefont {McNeile}},
  \bibinfo {author} {\bibfnamefont {E.}~\bibnamefont {Follana}}, \bibinfo
  {author} {\bibfnamefont {G.}~\bibnamefont {Lepage}}, \bibinfo {author}
  {\bibfnamefont {H.}~\bibnamefont {Na}},  \emph {et~al.} (\bibinfo
  {collaboration} {HPQCD Collaboration}),\ }\href {\doibase
  10.1103/PhysRevD.82.114504} {\bibfield  {journal} {\bibinfo  {journal}
  {Phys.Rev.}\ }\textbf {\bibinfo {volume} {D82}},\ \bibinfo {pages} {114504}
  (\bibinfo {year} {2010}{\natexlab{b}})},\ \Eprint
  {http://arxiv.org/abs/1008.4018} {arXiv:1008.4018 [hep-lat]} \BibitemShut
  {NoStop}%
\bibitem [{\citenamefont {Gregory}\ \emph {et~al.}(2011)\citenamefont
  {Gregory}, \citenamefont {Davies}, \citenamefont {Kendall}, \citenamefont
  {Koponen}, \citenamefont {Wong} \emph {et~al.}}]{gregory}%
  \BibitemOpen
  \bibfield  {author} {\bibinfo {author} {\bibfnamefont {E.~B.}\ \bibnamefont
  {Gregory}}, \bibinfo {author} {\bibfnamefont {C.~T.}\ \bibnamefont {Davies}},
  \bibinfo {author} {\bibfnamefont {I.~D.}\ \bibnamefont {Kendall}}, \bibinfo
  {author} {\bibfnamefont {J.}~\bibnamefont {Koponen}}, \bibinfo {author}
  {\bibfnamefont {K.}~\bibnamefont {Wong}},  \emph {et~al.} (\bibinfo
  {collaboration} {HPQCD Collaboration}),\ }\href {\doibase
  10.1103/PhysRevD.83.014506} {\bibfield  {journal} {\bibinfo  {journal}
  {Phys.Rev.}\ }\textbf {\bibinfo {volume} {D83}},\ \bibinfo {pages} {014506}
  (\bibinfo {year} {2011})},\ \Eprint {http://arxiv.org/abs/1010.3848}
  {arXiv:1010.3848 [hep-lat]} \BibitemShut {NoStop}%
\bibitem [{\citenamefont {Davies}\ \emph {et~al.}(1994)\citenamefont {Davies},
  \citenamefont {Hornbostel}, \citenamefont {Langnau}, \citenamefont {Lepage},
  \citenamefont {Lidsey} \emph {et~al.}}]{Davies:1994pz}%
  \BibitemOpen
  \bibfield  {author} {\bibinfo {author} {\bibfnamefont {C.}~\bibnamefont
  {Davies}}, \bibinfo {author} {\bibfnamefont {K.}~\bibnamefont {Hornbostel}},
  \bibinfo {author} {\bibfnamefont {A.}~\bibnamefont {Langnau}}, \bibinfo
  {author} {\bibfnamefont {G.}~\bibnamefont {Lepage}}, \bibinfo {author}
  {\bibfnamefont {A.}~\bibnamefont {Lidsey}},  \emph {et~al.},\ }\href
  {\doibase 10.1103/PhysRevLett.73.2654} {\bibfield  {journal} {\bibinfo
  {journal} {Phys.Rev.Lett.}\ }\textbf {\bibinfo {volume} {73}},\ \bibinfo
  {pages} {2654} (\bibinfo {year} {1994})},\ \Eprint
  {http://arxiv.org/abs/hep-lat/9404012} {arXiv:hep-lat/9404012 [hep-lat]}
  \BibitemShut {NoStop}%
\bibitem [{Note1()}]{Note1}%
  \BibitemOpen
  \bibinfo {note} {Dimensionless ratios of low-energy quantities are
  independent of the lattice-spacing scheme, and must be independent of~$m_c$
  by the decoupling theorem. This means that a scheme that makes any one
  low-energy quantity\protect \tmspace +\thinmuskip {.1667em}---\protect
  \tmspace +\thinmuskip {.1667em}for example, $w_0$\protect \tmspace
  +\thinmuskip {.1667em}---\protect \tmspace +\thinmuskip {.1667em}independent
  of~$m_c$ makes all other low-energy quantities independent of~$m_c$ as well,
  thereby preserving decoupling.}\BibitemShut {Stop}%
\end{thebibliography}%

\end{document}